\documentclass[reprint,showpacs,amsmath,amssymb,prd,]{revtex4-1}
\usepackage{dcolumn}
\usepackage{bm}
\usepackage[pdftex]{graphicx} 
\usepackage[utf8]{inputenc}
\usepackage{amsfonts}
\usepackage{graphicx}
\usepackage{subcaption}
\usepackage{color}
\usepackage{balance}
\usepackage{mathptmx}
\RequirePackage{flushend}
\usepackage{float}
\usepackage[normalem]{ulem}
\RequirePackage[colorlinks=true,citecolor=blue,urlcolor=blue,linkcolor=blue]{hyperref}
\DeclareMathAlphabet{\mathcal}{OMS}{cmsy}{m}{n}
\SetMathAlphabet{\mathcal}{bold}{OMS}{cmsy}{b}{n}
\begin{document}

\title{Echoes from a singularity}

\author{Avijit Chowdhury}
\email[Email: ]{ac13ip001@iiserkol.ac.in}
\affiliation{Indian Institute of Science Education and Research Kolkata, Mohanpur, Nadia, 741246, India}%

\author{Narayan Banerjee}
\email[Email: ]{narayan@iiserkol.ac.in}
\affiliation{Indian Institute of Science Education and Research Kolkata, Mohanpur, Nadia, 741246, India}%

\begin{abstract}
Though the cosmic censorship conjecture states that spacetime singularities must be hidden from an asymptotic observer by an event horizon, naked singularities can form as the end product of a gravitational collapse under suitable initial conditions, so the question of how to observationally distinguish such naked singularities from standard black hole spacetimes becomes important. In the present paper, we try to address this question by studying the ringdown profile of the Janis-Newman-Winicour (JNW) naked singularity under axial gravitational perturbation. The JNW spacetime has a surfacelike naked singularity that is sourced by a massless scalar field and reduces to the Schwarzschild solution in absence of the scalar field. We show that for low strength of the scalar field, the ringdown profile is dominated by echoes which mellows down as the strength of the field increases to yield characteristic quasinormal mode frequency of the JNW spacetime.
\end{abstract}
\maketitle
\section{Introduction}
Black holes are undoubtedly one of the most elegant constructs of general relativity that are characterized by a null surface, called the event horizon, which conceals a singularity within. Recent observations of gravitational waves by the LIGO-Virgo collaboration~\cite{ligo_PRL_2016, ligo_PRL_2016_1, ligo_PRL_2017, abbott_ApJL_2017, ligo_PRL_2017_1, ligo_PRL_2017_2, ligo_PRX_2019}, supplemented by the findings of the Event Horizon Telescope collaboration~\cite{eht_ApJ_2019, eht_ApJ_2019_1, zhu_ApJ_2018} have provided substantial evidence for the existence of these magnificent objects. 
However, probing the spacetime near the event horizon still remains an experimental challenge~\cite{eckart_FP_2017, abramowicz_AA_2002}. The black hole information paradox~\cite{unruh_RPP_2017} also heavily depends on the existence of the event horizon. Hence, one looks for exotic compact objects (ECOs) which are usually motivated by quantum gravity consideration. In an ECO model one assumes that quantum effects might intervene before the total collapse of a star to a black hole, forming a compact object devoid of an event horizon or a central singularity. Some examples are gravastars~\cite{mazur_arxiv_2001, mazur_NAS_2004}, boson stars~\cite{schunck_CQG_2003}, wormholes~\cite{morris_PRL_1988, damour_PRD_2007,  cardoso_PRL_2016}, fuzzballs~\cite{mathur_FP_2005} and others~\cite{barcelo_FP_2011, barcelo_CQG_2015, holdom_PRD_2017, konoplya_PRD_2019_1, posada_CQG_2019}. Another interesting possibility in classical gravity is the formation of a naked singularity, where the singularity is not cloaked by an event horizon and is visible to an asymptotic observer, violating the cosmic censorship conjecture~\cite{penrose_GERG_2002}. For gravitational collapse, the quantum considerations are generally towards an avoidance of a singularity~\cite{liu_PRD_2014, kiefer_PRD_2019, bojowald_PRL_2005}. Harada {\it et al.}~\cite{harada_PRD_2001} argued that although for a collapsing scenario of a quantized scalar field in a curved background the energy flux diverges near the cauchy horizon, this semiclassical  approach fails as the energy flux reaches the Planck scale, and there is no singularity. However, the study of gravitational collapse of massive matter clouds suggests that naked singularities can indeed form as an end-state under suitable initial conditions~\cite{choptuik_PRL_1993, joshi_PRD_1993, waugh_PRD_1988, christodoulou_CMP_1984, eardley_PRD_1979, ori_PRL_1987, giambo_GERG_2004, shapiro_PRL_1991, lake_PRD_1991, goswami_PRD_2007, joshi_PRD_2002, harada_PRD_1998, banerjee_PRD_2017, bhattacharya_PRD_2020}. An inevitable question that now arises is how to observationally distinguish such naked singularities from a black hole spacetime. The observational evidence for the existence of such naked singularities will not only invalidate the cosmic censorship conjecture, but will also have serious implications in the quantum gravity context.

In the electromagnetic spectrum, there are numerous observations pertaining to gravitational lensing~\cite{hioki_PRD_2009, yang_IJMPD_2016, takahashi_ApJ_2004, gyulchev_PRD_2008, virbhadra_PRD_2002, virbhadra_AA_1998, virbhadra_PRD_2008, bhattacharya_PRD_2020},  accretion disk~\cite{kovacs_PRD_2010, blaschke_PRD_2016, stuchlik_CQG_2010, bambi_PRD_2009, joshi_CQG_2011, pugliese_PRD_2011, bhattacharya_PRD_2020}, images and shadows~\cite{gyulchev_PRD_2019, chowdhury_PRD_2012, sau_PRD_2020} that highlight the difference in the properties of naked singularities and black hole spacetimes. In the gravitational wave spectrum such observations should also be possible, particularly, in the ringdown phase of a compact binary coalescence when the signal is dominated by the quasinormal modes. However, the templates for such signals are not known yet and further research is required for that.
The quasinormal modes (QNMs) are characterized by damped harmonic oscillations~\cite{kokkotas_LRR_1999,berti_CQG_2009} that can also be excited by linear perturbation of the compact object. In general, QNMs of different compact objects will be different and can be used to distinguish ECOs from black holes~\cite{debenedictis_CQG_2006, chirenti_CQG_2007, pani_PRD_2009, konoplya_PRD_2019_1, ferrari_PRD_2000, cardoso_PRL_2016, chirenti_PRD_2012, maggio_arxiv_2020}. The horizonless character of the ECOs may also be manifested by echoes in the ringdown profile~\cite{cardoso_PRD_2016, cardoso_NA_2017, mark_PRD_2017, konoplya_PRD_2019, chirenti_PRD_2020, churilova_CQG_2020, bronnikov_PRD_2020, roy_arxiv_2019, tsang_PRD_2018}.

In the present work we consider a static, spherically symmetric solution to the Einstein's equations with a naked singularity, known in literature~\cite{janis_PRL_1968, wyman_PRD_1981, virbhadra_IJMPA_1997} as the Janis-Newman-Winicour (JNW) naked singularity. The JNW naked singularity is  sourced by a massless scalar field. The JNW spacetime has a surfacelike naked singularity that reduces to the well known Schwarzschild solution in the absence of the scalar field. Optical properties of the JNW spacetime have been studied at great lengths~\cite{gyulchev_PRD_2008, virbhadra_PRD_2002, virbhadra_AA_1998, virbhadra_PRD_2008, gyulchev_PRD_2019, chowdhury_PRD_2012, sau_PRD_2020, jusufi_EPJC_2019, liu_JCAP_2018}. The JNW spacetime has been shown to be stable against scalar field perturbations~\cite{varadrajan_IJMPD_2012}. Scalar radiation from the JNW spacetime has also been studied in Refs.~\cite{liao_GERG_2014, dey_arxiv_2013}. Chirenti, Saa and Sk\'{a}kala~\cite{chirenti_PRD_2013}  studied the quasinormal modes of a test scalar field in the Wyman naked singularity~\cite{wyman_PRD_1981,virbhadra_IJMPA_1997} and showed the absence of asymptotically highly damped modes in the QNM spectrum.

In this paper, we aim to investigate the response of the JNW naked singularity-spacetime to axial (odd-parity) gravitational perturbation. We observe that even when the scalar field is weak, the signature of the difference between the spacetime due to a black hole and the naked singularity is quite distinctly elucidated  by the existence of echoes for the latter, but as the strength of the scalar field increases, the echoes align and the QNM structure of the JNW ringdown becomes prominent.

The paper is organized as follows. In Sec.~\ref{sec:2} we provide a brief review of the JNW naked singularity. Section~\ref{sec:3} discusses the gravitational perturbation of the JNW naked singularity-spacetime and obtains the master equation for axial gravitational perturbation. Section~\ref{sec:4} is dedicated to the time domain analysis of the perturbation equation and the evaluation of the associated quasinormal mode frequencies. Finally, in Sec.~\ref{sec:5} we conclude with a summary and discussion of the results that we arrived at. Throughout the paper, we employ units in which $G=c=1$.

\section{Review of the JNW naked singularity}\label{sec:2}
We start with an action in which the Einstein-Hilbert action is minimally coupled to a real massless scalar field $\Phi$,
\begin{equation}																				
S=\frac{1}{16 \pi}\int d^4 x \sqrt{-g}\left[\bar{R}- 8 \pi~\bar{g}^{\mu \nu} \partial_\mu \Phi \partial_\nu \Phi \right]~.			
\end{equation}																				
The field equations of the above theory,														
\begin{eqnarray}																				
\bar{R}_{\mu \nu} &=& 8 \pi \left( \partial_{\mu} \Phi \right) \left( \partial_{\nu} \Phi \right)~,\\
\square \Phi &=& 0,																			
\end{eqnarray}																				
 admit a static, spherically symmetric solution described by the line element,
 \begin{equation}\label{eq_metric}
ds^2=- f(r)dt^2+{f(r)}^{-1} dr^2+h^2(r)\left(d\theta^2 + \sin^2{\theta} d\phi^2 \right)~,
\end{equation}
where,
\begin{equation}\label{eq_metricfunction}
f(r)= \left(1-\frac{b}{r}\right)^{\nu} \mbox{and } h(r)=r \left(1-\frac{b}{r}\right)^\frac{1-\nu}{2}~.
\end{equation}
 This is the well known Janis-Newman-Winicour (JNW) spacetime~\cite{janis_PRL_1968, wyman_PRD_1981, virbhadra_IJMPA_1997}. The JNW spacetime is sourced by a scalar field,
 \begin{equation}
\Phi=\frac{q}{b\sqrt{4\pi}}\ln\left(1-\frac{b}{r}\right)~.
\end{equation}
The parameter $b$ is related to the scalar charge $q$ and the ADM mass $M$ as $b=2\sqrt{q^2+M^2}$ and $\nu=2M/b= M/\sqrt{q^2+M^2}$ lie in the range, $0 \leq \nu < 1$. For $q=0$ $\left(\nu=1 \right)$, the scalar field vanishes and one recovers the standard Schwarzschild metric. For an indefinitely high $q$,  $\nu$ reduces to 0. Thus, the parameter $\nu$ measures the deformation from the Schwarzschild spacetime. 
The JNW spacetime has a curvature singularity at $r=b$ for $0 < \nu < 1$. The absence of an event horizon makes the singularity globally naked. The spacetime also satisfies the weak energy condition~\cite{virbhadra_IJMPD_1997, virbhadra_IJMPA_1997}.  For $1/2<\nu< 1$, the JNW singularity lies within a photon sphere~\cite{claudel_JMP_2001, virbhadra_PRD_2002, virbhadra_PRD_2008} of radius
\begin{equation}
r_{ph}=\frac{b(1+2\nu)}{2}~,
\end{equation}
and the singularity is classified as weakly naked. However, for $0<\nu\leq1/2$ the singularity is no longer covered by a photon sphere and is classified as strongly naked. The JNW spacetime with weakly naked singularity is observationally characterized by a shadow and has lensing properties characteristic of the Schwarzschild black hole, whereas, for strongly naked singularity the lensing properties differ considerably from that of the Schwarzschild black hole~\cite{gyulchev_PRD_2008, virbhadra_PRD_2002, virbhadra_AA_1998, virbhadra_PRD_2008, gyulchev_PRD_2019, chowdhury_PRD_2012, sau_PRD_2020}.

In the present work, we will restrict ourselves only to the weakly naked singularity regime of the JNW spacetime. 

\section{Perturbation of the JNW spacetime}\label{sec:3}
We introduce small perturbations $h_{\mu \nu}$ to the background metric $\bar{g}_{\mu \nu}$ such that the resulting perturbed metric $g_{\mu \nu}$ becomes 
\begin{equation}
g_{\mu \nu}=\bar{g}_{\mu \nu} + h_{\mu \nu}~, \mbox{ where } \left|h_{\mu \nu}\right|/\left|\bar{g}_{\mu \nu}\right| \ll 1~.
\end{equation}
The perturbed metric gives rise to the perturbed  Christoffel symbols,
\begin{equation}
\Gamma^{\alpha}_{\mu \nu}=\bar{\Gamma}^{\alpha}_{\mu \nu}+\delta \Gamma^{\alpha}_{\mu \nu}~,
\end{equation}
where, $\bar{\Gamma}^{\alpha}_{\mu \nu}$ are the Christoffel symbols due to the unperturbed  metric and 
\begin{equation}\label{eq_deltaGamma}
\delta \Gamma^{\alpha}_{\mu \nu}= \frac{1}{2} \bar{g}^{\alpha \beta} \left(h_{\mu \beta, \nu} + h_{\nu \beta, \mu} - h_{\mu \nu, \beta} \right)~.
\end{equation}
This results in the perturbed Ricci tensor,
\begin{equation}
R_{\mu \nu} = \bar{R}_{\mu \nu}+\delta R_{\mu \nu}~,
\end{equation}
where
\begin{equation}\label{eq_deltaRicci}
\delta R_{\mu \nu}=\nabla_\nu \delta \Gamma^{\alpha}_{\mu \alpha}-\nabla_\alpha  \delta \Gamma^{\alpha}_{\mu \nu}~,
\end{equation}
and $\nabla_{\mu}$ is the covariant derivative with respect to the background metric $\bar{g}_{\mu \nu}$.

Due to spherical symmetry of the background spacetime, we can decompose the perturbations $\left( h_{\mu \nu} \right)$ into odd- $\left( h^{odd}_{\mu \nu} \right)$ and even-type $\left( h^{even}_{\mu \nu} \right)$ perturbations, based  on their parity under two dimensional rotation~\cite{regge_PRD_1957, zerilli_PRL_1970}. For excellent reviews, we refer to Refs.~\cite{berti_CQG_2009, rezzolla_review_2003}. In the present work, we will concentrate on the odd-parity or axial perturbation in which the perturbation $\delta \Phi$ of the background scalar field $\Phi$ does not contribute~\cite{kobayashi_PRD_2012}. Thus, the evolution of the axial perturbation is governed by the field equation,
 \begin{equation}\label{eq_deltaEinstein}
\delta R_{\mu \nu}=0~.
\end{equation}

The perturbation variables $h_{\mu \nu}$  can be expanded in a series of spherical harmonics. The components of the axial perturbation $\left( h^{odd}_{\mu \nu} \right)$ can be further simplified by utilizing the residual gauge freedom to choose a proper gauge. A preferred choice in this case is the ``Regge-Wheeler''  gauge~\cite{regge_PRD_1957} in which the axial perturbation is represented in terms of only two unknown functions $h_0 (t,r)$ and $h_1 (t,r)$,
\begin{equation}\label{eq_h}
h^{odd}_{\mu \nu}=\left(
\begin{array}{cccc}
 0& 0 & 0 & h_0(t,r)\\
 0 & 0 & 0 & h_1(t,r)\\
 0 & 0 & 0 & 0 \\
 h_0(t,r) & h_1(t,r) & 0 & 0 \\
\end{array}
\right) \sin\theta \partial_{\theta} P_\ell(\cos \theta) e^{i m \phi}~,
\end{equation}
where $ P_\ell(\cos \theta)$ is the Legendre polynomial of order $\ell$ and $m$ is the azimuthal harmonic index.

\begin{figure}[!htbp]
\begin{center}
  \includegraphics[scale=1]{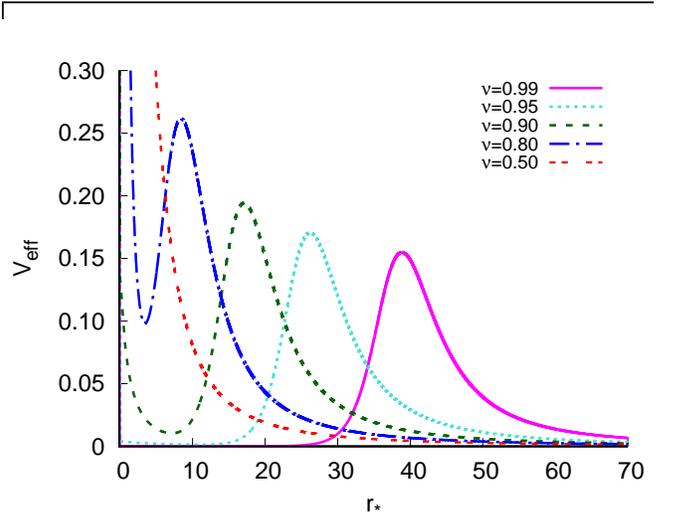}
\caption{Plot of the effective potential with $r_*$ for different values of $\nu$ with $b=2$ and $\ell=2$.} 
\label{fig_potential_trts}
\end{center}
\end{figure}

$\delta R_{\mu \nu}$ has ten components, out of which only the $(t,\phi)$, $(r,\phi)$ and $(\theta, \phi)$ components are nonzero. We explicitly write the $(t,\phi)$, $(r,\phi)$ and $(\theta, \phi)$ components of Eq.~\eqref{eq_deltaEinstein} in a simplified form as,
\begin{widetext}
\begin{equation}\label{eq_deltaEinstein03}
\begin{split}
\delta R_{t\phi}=&-\frac{\left(1-\frac{b}{r}\right)^{\nu} h_0(t,r) \left(b^2 \nu  (\nu +1)-b r \left(2 \nu +\ell ^2+\ell \right)+r^2 \ell  (\ell +1)\right)}{2 r^2 (b-r)^2}-\\
&\frac{1}{2 r (b-r)}\left(1-\frac{b}{r}\right)^{\nu }\left(r (b-r) \partial^2_r h_0(t,r)\right.
-\left.(b \nu +b-2 r) \partial_t h_1(t,r)+r (r-b) \partial_t \partial_r h_1(t,r)\right)=0~,
\end{split}
\end{equation}
\begin{equation}\label{eq_deltaEinstein13}
\begin{split}
\delta R_{r \phi}=&\frac{1}{2} r^{\nu -1} (r-b)^{-\nu -1} \left(r (b-r) \left(\partial_t \partial_r h_0(t,r)-\partial^2_t h_1(t,r)\right)-\right.\\
&\left.(b \nu +b-2 r) \partial_t h_0(t,r)\right)-\frac{\left(\ell ^2+\ell -2\right) \left(1-\frac{b}{r}\right)^{\nu } h_1(t,r)}{2 r (b-r)}=0~,
\end{split}
\end{equation}
\begin{equation}\label{eq_deltaEinstein23}
\delta R_{\theta \phi}=\frac{1}{2} \left(1-\frac{b}{r}\right)^{-\nu } \left(\partial_t h_0(t,r)-\left(1-\frac{b}{r}\right)^{2 \nu } \partial_r h_1(t,r)\right)+\frac{b \nu  r^{-\nu -1} (r-b)^{\nu } h_1(t,r)}{2 b-2 r}=0~.
\end{equation}
\end{widetext}
Substituting $\partial_t h_0(t,r)$ from Eq.~\eqref{eq_deltaEinstein23} to Eq.~\eqref{eq_deltaEinstein13} and defining $\Psi(t,r)=\frac{h_1(t,r)}{r}(1-b/r)^{\frac{3\nu -1}{2}}$, we obtain Eq.~\eqref{eq_deltaEinstein13} in a Schr\"{o}dinger-like form,
\begin{equation}\label{eq_RW}
\frac{\partial^2 }{\partial t ^2}\Psi(t,r)-\frac{\partial^2}{\partial r_* ^2}\Psi(t,r)+V_{eff}(r)  \Psi(t,r)=0~,
\end{equation}
where
\begin{equation}
\begin{split}
V_{eff}(r)=&\frac{1}{4} r^{-2 (\nu +1)} (r-b)^{2 (\nu -1)} \left(3 b^2 (\nu +1)^2-\right.\\
&\left.4 b r \left(3 \nu +\ell ^2+\ell \right)+4 r^2 \ell  (\ell +1)\right)
\end{split}
\end{equation}
is the effective potential.
The coordinate $r_*$ is known as the tortoise coordinate and is defined by the relation, 
\begin{equation}\label{eq_tortoise}
\frac{dr_*}{dr}=\left(1-\frac{b}{r}\right)^{-\nu}~.
\end{equation}
For  $\nu \in \left(0,1\right)$, the tortoise coordinate maps the singularity at $r=b$ to $r_*=0$.
The effective potential vanishes as $r_*\rightarrow \infty$, whereas, near the singularity it  rises to an infinite wall,
\begin{equation}
V_{eff}( r\rightarrow b )\rightarrow \infty \quad \mbox{for} \quad 0 < \nu < 1~.
\end{equation}
The effective potential as a function of the tortoise coordinate $r_*$ for different values of the parameter $\nu$ is depicted in Fig.~\ref{fig_potential_trts}. For numerical simplicity the origin of the tortoise coordinate in Fig.~\ref{fig_potential_trts} has been shifted from $r=b$ to $r=b+\epsilon ~ \left( \epsilon<< 1\right)$. We have chosen $b=2$, i.e., $q^2 + M^2 =1 $ for a better control over the numerical work. This implies we actually work on the basis of the relative strength of the scalar charge $q$ to the mass $M$, for $\nu = 1$, $q=0$ and the system reduces to a Schwarzschild geometry, whereas for $\nu= \frac{1}{2}$, $\frac{q}{M}$ is as high as $\sqrt{3}$. 
Away from the singularity, for $\nu$ in the range $\left(1/2, 1\right)$, the effective potential is characterized by a peak. As $\nu$ decreases from $\nu\approx 1$ to $\nu\approx 1/2$, the height of the potential peak increases and it moves closer to the $r_*=0$ surface. For $0<\nu\leq 1/2$, i.e., in the strongly naked singularity regime, the peak vanishes and the potential profile is solely characterized by a potential wall, gradually rising to infinity at the singularity.
\section{Time-domain profile and quasinormal modes}\label{sec:4}
In order to study the time evolution of the perturbation, we rewrite the wave equation~\eqref{eq_RW}  in terms of the light-cone (null) coordinates, $u=t-r_*$ and $v=t+r_*$, as,
\begin{equation}\label{eq_RW_uv}
4\frac{\partial ^2}{\partial u \partial v}\Psi(u,v)+V_{eff}(u,v)\Psi(u,v)=0~.
\end{equation}
The appropriate discretization  scheme to integrate Eq.~\eqref{eq_RW_uv} as proposed by Gundlach, Price and Pullin~\cite{gundlach_PRD_1994} is,
\begin{equation}\label{eq_TD}
\begin{split}
\Psi(N)=&\Psi(W)+\Psi(E)-\Psi(S)\\
&-\Delta^2 \frac{V_{eff}(W)\Psi(W)+V_{eff}(E)\Psi(E)}{8}+\mathcal{O}\left(\Delta^4\right)~,
\end{split}
\end{equation}
where we have used the following designations for the points in the $u-v$ plane with step-size $\Delta$: $N=(u+\Delta,v+\Delta)$, $W=(u+\Delta,v)$, $E=(u,v+\Delta)$ and $S=(u,v)$.
In the linear regime, the eigenfrequencies of the JNW spacetime are not sensitive to the choice of the initial condition, so, we model the initial perturbation by a Gaussian pulse of width $\sigma$ centred around $v=v_c$,
\begin{equation}
\Psi (u=0,v)=e^{-\frac{\left( v-v_c \right)^2}{2 \sigma^2}}~,
\end{equation}
and assume the perturbation to vanish at $r_*=0$~,
\begin{equation}\label{eq_bdry}
\Psi (r_* =0,t)=\Psi(u=v,v)= 0, \quad \forall t~.
\end{equation}
The choice of the boundary condition, given in Eq.~\eqref{eq_bdry} deserves some attention. It should be noted that the JNW spacetime fails to be globally hyperbolic due to the naked singularity at $r=b$. A general prescription to define sensible dynamics of the perturbation field (in the linear regime) in such static, nonglobally hyperbolic spacetimes has been suggested by Wald ~\cite{wald_JMP_1980} (see also Refs.~\cite{horowitz_PRD_1995, ishibashi_PRD_1999, ishibashi_CQG_2003, helliwell_GERG_2003, ishibashi_CQG_2004, gibbons_PTP_2005, cardoso_PRD_2006}). One starts by defining an operator $A$ denoting the spatial (derivative) part of Eq.~\eqref{eq_RW},
\begin{equation}
A=-\frac{d^2}{d r_* ^2}+V_{eff}~.
\end{equation}
The operator $A$ acts on the Hilbert space $\mathcal{H}=L^2\left(r_* ,dr_*\right)$, of square integrable functions on the static hypersurface $\Sigma$, orthogonal to the unit timelike vector. The existence of a unique self adjoint extension $A_E$ of $A$ guarantees unitary dynamical evolution of the perturbation field, which, in our case, corresponds to choosing the appropriate boundary condition at the singularity.

As we approach the singularity, we can write $f(r)\approx b^{-\nu}\left({r-b}\right)^\nu$, $\quad h(r)\approx b^\frac{1+\nu}{2} (r-b)^\frac{1-\nu}{2}$ and the tortoise coordinate, $r_*\approx \frac{b^\nu}{1-\nu}(r-b)^{1-\nu}$. Close to the singularity, the effective potential reduces to,
\begin{equation}
V_{eff}\left(r_*\right)\approx\frac{3}{4 r_*^2}+\mathcal{O}\left(\frac{1}{r_* ^{\gamma}}\right)~,
\end{equation}
where $\gamma=-\frac{2\nu-1}{1-\nu} <2$~. 
Assuming, 
\begin{equation}
\Psi\left(t,r_*\right)= e^{-i\omega t}\Psi\left(r_*\right)~,
\end{equation}
we get
\begin{equation}\label{eq_operator}
A\Psi \equiv-\frac{d^2\Psi}{d r_* ^2}+V_{eff}\Psi=\omega^2\Psi~,
\end{equation}
which at the leading order reduces to
\begin{equation}\label{eq_BC_sing}
-\frac{d^2\Psi}{d r_*^2}+\frac{3}{4 r_* ^2}\Psi=\omega^2 \Psi~,\quad \mbox{as} \quad r_*\rightarrow 0~.
\end{equation}
The general solution to Eq.~\eqref{eq_BC_sing} is given by,
\begin{equation}\label{eq_gen_sol}
\Psi \sim \mathcal{C}_1\left({r_*^{-1/2}} + \cdots\right)+\mathcal{C}_2 \left( r_*^{3/2} + \cdots \right)~,\quad \mbox{as}\quad r_*\rightarrow 0.
\end{equation}
Equation~\eqref{eq_gen_sol} suggests that for $\Psi$ to be normalizable, we must have $\mathcal{C}_1=0$ i.e., $\Psi$ must satisfy 
\begin{equation}\label{eq_slfadjext}
r_*^{1/2}\Psi \mid_{r_{*} =0}=0~.
\end{equation}
With the boundary condition \eqref{eq_bdry}, one can show that the differential operator $A$ of \eqref{eq_operator} has a unique Friedrichs extension. Thus the choice of the boundary condition is consistent following Wald's seminal work~\cite{wald_JMP_1980}. A very brief description of the method is given in Appendix~\ref{appendix1}.

It is important to mention that for scalar field propagation in the Wyman spacetime, which is actually identical with the JNW metric as shown by  Virbhadra~\cite{virbhadra_IJMPA_1997}, the time translation operator also has a unique self-adjoint extension~\cite{ishibashi_PRD_1999}. Chirenti, Saa and Sk\'{a}kala~\cite{chirenti_PRD_2013}, have also shown the uniqueness of the time evolution  of scalar fields in the Wyman spacetime.

\begin{figure*}[!htb]
\begin{subfigure}{0.33\textwidth}
  \centering
  \includegraphics[width=\linewidth]{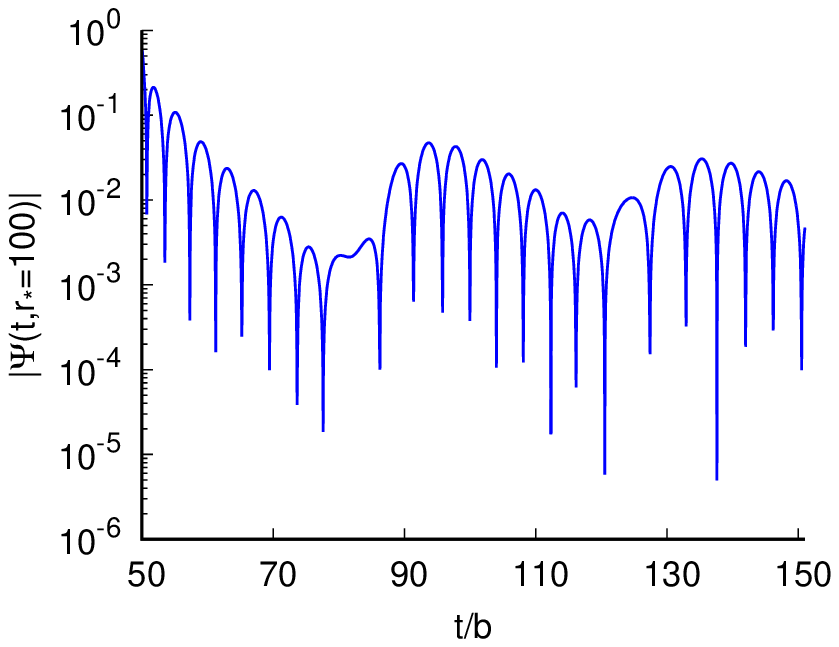}
\end{subfigure}%
\begin{subfigure}{.33\textwidth}
  \centering
  \includegraphics[width=\linewidth]{{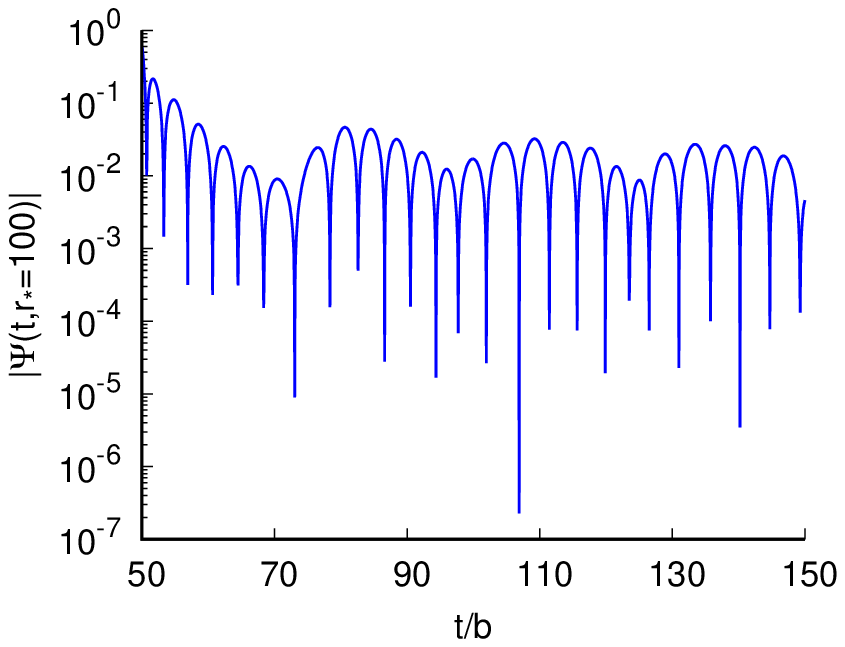}}
\end{subfigure}
\begin{subfigure}{0.33\textwidth}
  \centering
  \includegraphics[width=\linewidth]{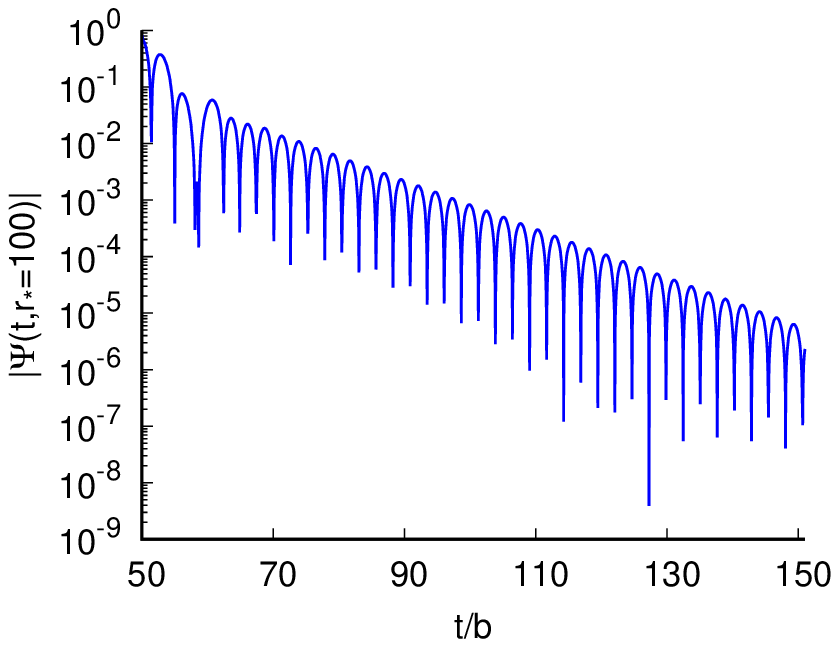}
\end{subfigure}\\
\begin{subfigure}{0.33\textwidth}
  \centering
  \includegraphics[width=\linewidth]{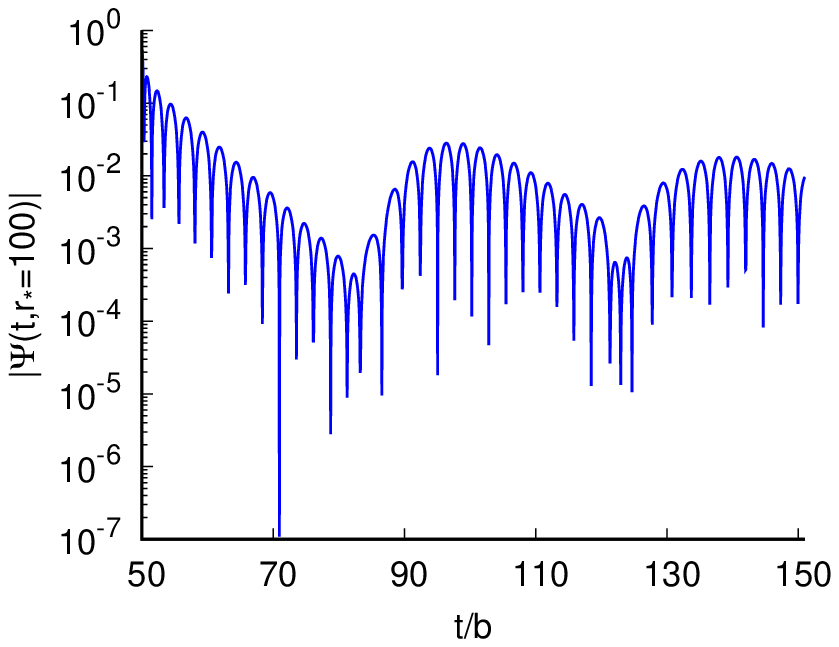}
\end{subfigure}%
\begin{subfigure}{.33\textwidth}
  \centering
  \includegraphics[width=\linewidth]{{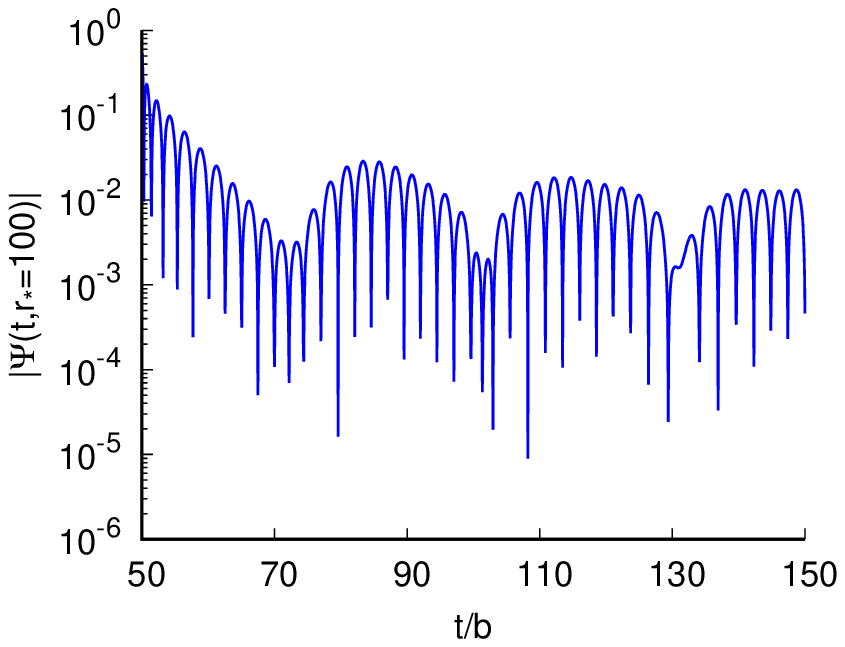}}
\end{subfigure}
\begin{subfigure}{0.33\textwidth}
  \centering
  \includegraphics[width=\linewidth]{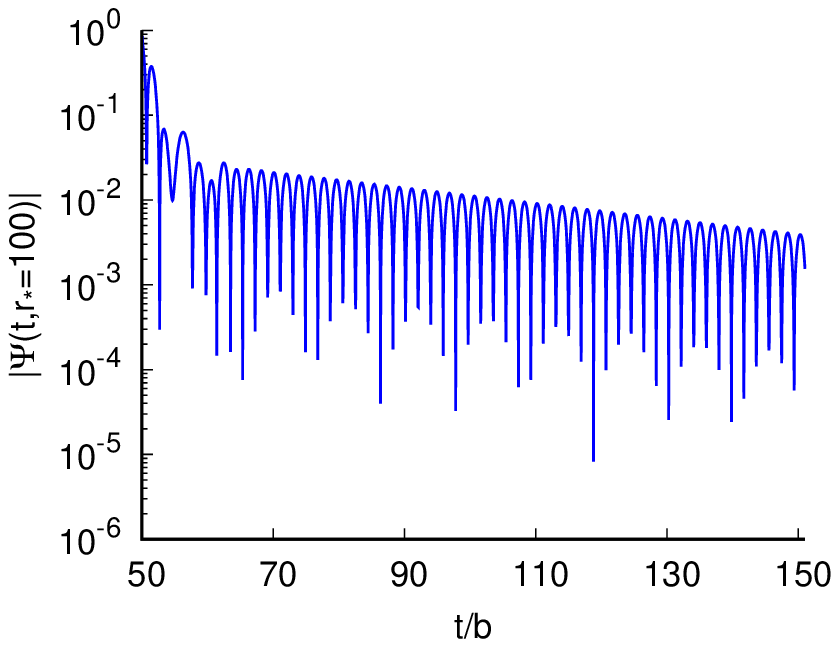}
   \end{subfigure}
 \caption{Semilogarithmic plot of the time-domain profiles for axial gravitational perturbation of the JNW spacetime as extracted at $r_*=$100 with $b=$2 for $\nu=$0.99, 0.95, 0.75 (from left to right) and $\ell=$2, 3 (from top to bottom). The horizontal axis has been scaled by $b$ to make it dimensionless.}
\label{fig:td_echo}
\end{figure*}

Using the integration scheme given in Eq.~\eqref{eq_TD}, we study the time evolution of the field $\Psi$ along a line of constant $r_*$. Figure~\ref{fig:td_echo}  shows the time evolution of axial perturbation of the JNW space time for different values of the parameter $\nu$. The qualitative features of the ringdown profile are clearly described by the plots.
We find that close to the Schwarzschild limit $\left( \nu= 1\right)$, when scalar charge $q$ is small, the response of the JNW spacetime is dominated by damped  harmonic oscillations, which soon give way to distinctive echoes. The echoes cannot be characterized by a single dominant frequency. However, with the decrease in $\nu$, as the peak of the effective potential moves closer to the wall, the echoes become less prominent and finally, the enveloping oscillation of the echoes align to yield  characteristic frequencies of the JNW spacetime.

To extract the characteristic  quasinormal mode frequencies from the time-domain profile we use the Prony method of fitting the data via superposition of damped exponentials with some excitation factors~\cite{konoplya_review_2011, berti_PRD_2007},
\begin{equation}\label{eq_prony1}
\Psi(t)\simeq\sum_{j=1}^{p}C_j e^{-i \omega_j t}~.
\end{equation}
We provide a brief description of the Prony method in Appendix~\ref{appendix2}. It deserves mention that the usual WKB method of evaluating the quasinormal mode frequencies~\cite{schutz_APJ_1985, iyer_PRD_1987, matyjasek_PRD_2017, konoplya_CQG_2019, konoplya_PRD_2003, blome_PLA_1984} does not work in the present case. The WKB method is applicable only when the potential function, $U(x,\omega)=V_{eff}(r)-\omega^2$ has two turning points, which in general is not the case for the JNW spacetime due to the presence of the infinite potential wall.
\begin{table}[!tb]
\caption{Fundamental quasinormal mode frequencies (in units of $b$) for axial gravitational perturbation of the JNW spacetime for $\ell=2,3$. The data is indicative of the temporal sequence, QNMs followed by echoes for small $\nu$ and initial outburst (I.O.) followed by QNMs for large $\nu$.}
\label{table:1}
\begin{tabular}{ccccc}
	\hline \hline 
	\rule[-1ex]{0pt}{2.5ex} ~$\nu$ ~&~ $\omega~(\ell=2)$ ~&~ $\omega~(\ell=3)$  \\
	\hline
	\rule[-1ex]{0pt}{2.5ex} ~$1.00$ ~&~ $0.3730 - 0.0891 i$ ~&~ $0.5993 - 0.0927 i$ \\ 
	\rule[-1ex]{0pt}{2.5ex} ~$0.99$ ~&~  $0.4066 - 0.1001 i$, echo ~&~ $0.6070 - 0.0940 i$, echo \\  
	\rule[-1ex]{0pt}{2.5ex} ~$0.95$ ~&~ $0.4202 - 0.1013 i$, echo ~&~ $0.6362 - 0.0961 i$, echo \\ 
	\rule[-1ex]{0pt}{2.5ex} ~$0.86$ ~&~  I.O.~, $0.3404 - 0.0004 i$ ~&~ $0.7290 - 0.0968 i$, echo \\ 
	\rule[-1ex]{0pt}{2.5ex} ~$0.83$ ~&~ I.O.~, $0.4168 - 0.0033 i$ ~&~ I.O.~, $0.5154 - 0.00003 i$ \\  
	\rule[-1ex]{0pt}{2.5ex} ~$0.80$ ~&~ I.O.~, $0.4903 - 0.0122 i$ ~&~I.O.~, $0.6324 - 0.0004 i$ \\  
	\rule[-1ex]{0pt}{2.5ex} ~$0.78$ ~&~ I.O.~, $0.5371-0.0235 i$ ~&~I.O.~, $0.7103 - 0.0019 i$ \\  
	\rule[-1ex]{0pt}{2.5ex} ~$0.75$ ~&~ I.O.~, $0.6044 - 0.0492 i$ ~&~I.O.~, $0.8229 - 0.0105 i$ \\ 
	\rule[-1ex]{0pt}{2.5ex} ~$0.73$ ~&~ I.O.~, $0.6471 - 0.0725 i$ ~&~I.O.~, $0.8947 - 0.0233 i$ \\  
	\hline \hline 
\end{tabular} 
\end{table}

Table~\ref{table:1} shows the fundamental quasinormal mode frequencies of the weakly naked JNW spacetime. The quasinormal modes for the Schwarzschild case ($\nu =1$) correspond to the standard black hole - boundary condition of completely ingoing waves at the horizon and completely outgoing waves at spatial infinity. The quasinormal mode frequency for $\nu=1$ matches with that obtained in Refs.~\cite{chandra_PRSL_1975, iyer_PRD_1987, shu_PLB_2005}. $\omega$ corresponding to $\nu=0.99, 0.95$ for $\ell=2$ and $\nu=0.99, 0.95, 0.86$ for $\ell=3$ represent the dominant quasinormal mode frequency of the initial oscillatory falloff, characteristic of the potential peak at $r>b$, which later gives rise to echo signals. We see from Table~\ref{table:1} that for a given multipole index, $\ell$, once the scalar charge becomes significantly large and the echoes align ($\nu=0.86 -0.73$ for $\ell=2$ and $\nu=0.83-0.73$ for $\ell=3$), both the oscillation frequency and the damping rate of the fundamental quasinormal mode (the real and imaginary parts of the QNM frequency respectively) increase with the decrease of $\nu$. With increasing values of the multipole index the echoes persist for even smaller values of $\nu$.

An important parameter in the analysis of ringdown signal is the quality factor which is defined as the ratio of the real and imaginary parts of the quasinormal mode frequency
\begin{equation}
Q\sim\frac{Real(\omega)}{\left| Im\left(\omega \right)\right|}~.
\end{equation}
The quality factor for the quasinormal modes of the Schwarzschild black hole and that of the JNW spacetime is shown in Table~\ref{table:2}.
\begin{table}
\caption{Quality factor for the fundamental quasinormal mode frequencies for axial gravitational perturbation of the JNW spacetime with $\ell=2,3$. }
\label{table:2}
\begin{tabular}{cccc}
	\hline \hline 
	\rule[-1ex]{0pt}{2.5ex} $\nu$ & $Q~(\ell=2)$ & $Q~(\ell=3)$ \\ 
	\hline
	\rule[-1ex]{0pt}{2.5ex} $1.00$ & $ 4.186$ & $6.465$ \\
	\rule[-1ex]{0pt}{2.5ex} $0.99$ & $4.062$ & $6.457$ \\ 
	\rule[-1ex]{0pt}{2.5ex} $0.95$ & $ 4.148$ & $6.620$ \\ 
	\rule[-1ex]{0pt}{2.5ex} $0.86$ & $851$ & $7.531$ \\ 
	\rule[-1ex]{0pt}{2.5ex} $0.83$ & $126.303$ & $17180$ \\ 
	\rule[-1ex]{0pt}{2.5ex} $0.80$ & $40.189$ & $1581$ \\ 
	\rule[-1ex]{0pt}{2.5ex} $0.78$ & $22.855$ & $373.842$ \\ 
	\rule[-1ex]{0pt}{2.5ex} $0.75$ & $12.285$ & $78.371$ \\
	\rule[-1ex]{0pt}{2.5ex} $0.73$ & $8.926$ & $38.399$ \\
	\hline \hline 
\end{tabular} 
\end{table}
We note from Table~\ref{table:2} that (starting from $\nu=0.99$) for a given multipole index, the quality factor increases with the decrease in $\nu$, reaches a maximum as the echoes in the ring-down profile aligns and then falls off with further decrease in $\nu$.
\section{Conclusion}\label{sec:5}
The detection of gravitational waves from compact binary coalescence has opened a new window of opportunity to probe the strong field regime of gravity and, hence to test the existence of exotic compact objects. One such horizonless compact object is a naked singularity, which can form as a result of gravitational collapse under suitable initial conditions.

In the present work, we studied the ringdown profile of the Janis-Newman-Winicour  naked singularity-spacetime~\cite{janis_PRL_1968, wyman_PRD_1981, virbhadra_IJMPA_1997} for an axial gravitational perturbation. We specifically investigated the weakly naked singularity regime where the spacetime is characterized by a photon sphere. At the singularity at $r=b$, the effective potential becomes infinite. For the QNM pertaining to a black hole one has an advantage of picking up boundary conditions elegantly, pure incoming waves at the horizon and pure outgoing waves at large distances. A purely incoming or outgoing mode is not suitable in the present case. For unique time evolution of the perturbation field, we choose null Dirichlet condition at the singularity, consistent with the work of Chirenti, Saa and Sk\'{a}kala~\cite{chirenti_PRD_2013}. 

Near the Schwarzschild limit, the initial response of the spacetime is characterized by damped oscillations, reminiscent of the potential peak at $r>b$. At later times, these damped oscillations give rise to distinct echoes. However, as the parameter $\nu$ is decreased, the echoes die down and characteristic quasinormal modes emerge. In the extreme right plots of Fig.~\ref{fig:td_echo}, both up and down, when $\nu = 0.75$, corresponding to $\frac{q}{M} = \frac{\sqrt{7}}{3}$, there are no echoes. Certainly the QNMs are different from the black holes. As the echoes align, the QNM spectrum is dominated by modes with very low damping rate, hence, high quality factor (as evident from Tables~\ref{table:1} and \ref{table:2} for $\nu=0.86$ for $\ell=2$ and $\nu=0.83$ for $\ell=3$). Thus, the JNW spacetime in this range is an excellent oscillator. With further decrease in $\nu$, both the oscillation frequency and the damping rate of the quasinormal modes increase which in turn reduces the quality factor of the oscillation. It is also important to note that in the entire analysis we do not observe any unstable quasinormal modes with frequency having positive imaginary part. 

The existence of echoes in the ringdown signal of the JNW naked singularity is a novel result which categorically differentiates it from a black hole. It deserves mention that Chirenti, Saa and Sk\'{a}kala~\cite{chirenti_PRD_2013} showed the absence of asymptotically highly damped modes in the Wyman naked singularity~\cite{wyman_PRD_1981, virbhadra_IJMPA_1997}, but their work was based on the perturbation of a test scalar field whereas the present work is based on the tensor perturbation of the metric.
Further investigations may be able to compare the ringdown profile of the JNW spacetime with that of other exotic compact objects such as wormholes which produce similar echoes in the ringdown phase~\cite{maggio_arxiv_2020, cardoso_PRD_2016, cardoso_NA_2017, mark_PRD_2017, konoplya_PRD_2019, chirenti_PRD_2020, churilova_CQG_2020, bronnikov_PRD_2020, roy_arxiv_2019}. Low power echoes may also be produced by black holes, provided, one considers a dramatic deviation from general relativity or effective field theory or both~\cite{damico_PRD_2019}. We also plan to extend our analysis to polar perturbations (even parity) as well. In this regard it is also  important to note that the perturbation of the scalar field $\Phi$, which does not contribute in the present odd-parity case, will give rise to breathing modes~\cite{kar_PRD_2018} in the QNM spectrum, which results purely from $\delta \Phi$.

\bigskip
\begin{acknowledgements}
A.C. wishes to thank Prof. Rajesh Kumble Nayak and Abhishek Majhi for useful discussions.
\end{acknowledgements}
\appendix
\section{Friedrichs extension}\label{appendix1}
Whether an operator $A$ admits a self-adjoint extension or not depends on if the so-called the von Neumann criterion is satisfied or not. For a given operator $A$, one looks for 
the solution for $\phi$ from the equation $A\phi=\pm i \phi$. The dimension of the solution space is called the deficiency index $n$. One will have two values for $n$, namely $n_{\pm}$, corresponding to the two eigenvalues $\pm i$. For $n_{\pm} = 0$, there is a unique Friedrichs extension. For nonzero values of the deficiency indices, a self-adjoint extension is still  possible if $n_+ = n_-$, although, the uniqueness of the extension is not guaranteed in this case.

The relevant operator in the present work, $A$, as given in Eq.~\eqref{eq_operator}, satisfies $n_{\pm} = 0$. For an elegant discussion on self-adjoint extension, we refer to Ref.~\cite{reed_simon_1975} (see also Ref. ~\cite{sridip_JMP_2016}). 

\section{Prony method for extracting QNM frequencies}\label{appendix2}
The fitting of the time-profile data with a sum of damped exponentials can, in general, be performed using standard nonlinear least-squares techniques. However, such fitting is not very accurate~\cite{berti_PRD_2007}. To reduce the squared error over the data, one needs to numerically solve highly nonlinear equations involving sums of the powers of damping coefficients. Such solutions depend critically on the initial guesses for the parameters~\cite{marple_book}. Though iterative methods such as Newton's method or gradient descent algorithms can be used to address the optimization problems, they are computationally very  expensive~\cite{mcdonough_IEEE_1968,evans_IEEEAES_1973}. This led to the development of new class of fitting techniques relying on the method originally developed by Prony in 1795~\cite{prony}.

We start with the assumption that the ringdown begins at $t_0=0$ and continues until $t=N h$, where $N\geq 2p-1$, such that for each $n\in\left(0,N\right)$
\begin{equation}
x_n\equiv \Psi(n h)=\sum_{j=1}^{p}C_j e^{-i \omega_j n h}=\sum_{j=1}^{p} C_j z_j^n ~.
\end{equation}
The Prony method provides an ingenious way to determine $z_j$ from the profile data $x_n$, thereby allowing to calculate the quasinormal mode frequencies $\omega_j$. We define a polynomial $\tilde{A}(z)$ of degree $p$ as
\begin{equation}\label{eq_prony2}
\tilde{A}(z)=\displaystyle\prod_{j=1}^p(z-z_j)=\sum_{k=0}^p \alpha_k z^{p-k}, \quad \alpha_0=1~.
\end{equation}
and consider the summation,
\begin{equation}\label{eq_prony3}
\sum_{k=0}^p \alpha_k x_{n-k}=\sum_{k=0}^p \alpha_k \sum_{j=1}^p C_j z_j^{n-k}= \sum_{j=1}^p C_j z_j^{n-p}\sum_{k=0}^p \alpha_k z_j^{p-k}~.
\end{equation}
Using Eq.~\eqref{eq_prony2} in Eq.~\eqref{eq_prony3} we get,
\begin{equation}
\sum_{k=0}^p \alpha_k x_{n-k}=\sum_{j=1}^p C_j z_j^{n-p} \tilde{A}(z)=0
\end{equation}
which can be rewritten as,
\begin{equation}\label{eq_prony4}
\sum_{k=1}^p \alpha_k x_{n-k}=-x_n~.
\end{equation}
Substituting $n=p,\cdots,N$ in Eq.~\eqref{eq_prony4}, we get $N-P+1 \geq p$ linear equations, which can be written in the matrix form,
\begin{equation}\label{eq_prony5}
\mathbf{X \alpha} =-\mathbf{x}
\end{equation}
where
\begin{equation}
\mathbf{X}=\begin{pmatrix}
x_{p-1} & x_{p-2} & \cdots & x_0 \\
x_{p} & x_{p-1} & \cdots & x_1 \\
\vdots  & \vdots  & \ddots & \vdots  \\
x_{N-1} & x_{N-2} & \cdots & x_{N-p} 
\end{pmatrix}, \mathbf{\alpha}=\begin{pmatrix}
\alpha_1\\
\alpha_2\\
\vdots\\
\alpha_p
\end{pmatrix}, \mathbf{x}=\begin{pmatrix}
x_p\\
x_{p+1}\\
\vdots\\
x_N
\end{pmatrix}~.
\end{equation}
Equation~\eqref{eq_prony5} can be solved in the least-squares sense to determine the unknown coefficient matrix $\mathbf{\alpha}$,
\begin{equation}
\mathbf{\alpha}=-\left(\mathbf{X^\dagger X} \right)^{-1} \mathbf{X^\dagger x}~,
\end{equation}
where $\mathbf{X^\dagger}$ is the Hermitian transpose of $\mathbf{X}$.
Having determined the coefficients $\alpha_k$ and hence the roots $z_j$ of the polynomial $\tilde{A}(z)$ we obtain the quasinormal mode frequencies,
\begin{equation}
\omega_j=\frac{i}{h}\ln\left(z_j\right)~.
\end{equation}
\bibliography{ref.bib}

\begin{thebibliography}{119}%
\makeatletter
\providecommand \@ifxundefined [1]{%
 \@ifx{#1\undefined}
}%
\providecommand \@ifnum [1]{%
 \ifnum #1\expandafter \@firstoftwo
 \else \expandafter \@secondoftwo
 \fi
}%
\providecommand \@ifx [1]{%
 \ifx #1\expandafter \@firstoftwo
 \else \expandafter \@secondoftwo
 \fi
}%
\providecommand \natexlab [1]{#1}%
\providecommand \enquote  [1]{``#1''}%
\providecommand \bibnamefont  [1]{#1}%
\providecommand \bibfnamefont [1]{#1}%
\providecommand \citenamefont [1]{#1}%
\providecommand \href@noop [0]{\@secondoftwo}%
\providecommand \href [0]{\begingroup \@sanitize@url \@href}%
\providecommand \@href[1]{\@@startlink{#1}\@@href}%
\providecommand \@@href[1]{\endgroup#1\@@endlink}%
\providecommand \@sanitize@url [0]{\catcode `\\12\catcode `\$12\catcode
  `\&12\catcode `\#12\catcode `\^12\catcode `\_12\catcode `\%12\relax}%
\providecommand \@@startlink[1]{}%
\providecommand \@@endlink[0]{}%
\providecommand \url  [0]{\begingroup\@sanitize@url \@url }%
\providecommand \@url [1]{\endgroup\@href {#1}{\urlprefix }}%
\providecommand \urlprefix  [0]{URL }%
\providecommand \Eprint [0]{\href }%
\providecommand \doibase [0]{http://dx.doi.org/}%
\providecommand \selectlanguage [0]{\@gobble}%
\providecommand \bibinfo  [0]{\@secondoftwo}%
\providecommand \bibfield  [0]{\@secondoftwo}%
\providecommand \translation [1]{[#1]}%
\providecommand \BibitemOpen [0]{}%
\providecommand \bibitemStop [0]{}%
\providecommand \bibitemNoStop [0]{.\EOS\space}%
\providecommand \EOS [0]{\spacefactor3000\relax}%
\providecommand \BibitemShut  [1]{\csname bibitem#1\endcsname}%
\let\auto@bib@innerbib\@empty
\bibitem [{\citenamefont {Abbott}\ \emph
  {et~al.}(2016{\natexlab{a}})\citenamefont {Abbott} \emph
  {et~al.}}]{ligo_PRL_2016}%
  \BibitemOpen
  \bibfield  {author} {\bibinfo {author} {\bibfnamefont {B.~P.}\ \bibnamefont
  {Abbott}} \emph {et~al.} (\bibinfo {collaboration} {LIGO Scientific,
  Virgo}),\ }\href {\doibase 10.1103/PhysRevLett.116.241103} {\bibfield
  {journal} {\bibinfo  {journal} {Phys. Rev. Lett.}\ }\textbf {\bibinfo
  {volume} {116}},\ \bibinfo {pages} {241103} (\bibinfo {year}
  {2016}{\natexlab{a}})}\BibitemShut {NoStop}%
\bibitem [{\citenamefont {Abbott}\ \emph
  {et~al.}(2016{\natexlab{b}})\citenamefont {Abbott} \emph
  {et~al.}}]{ligo_PRL_2016_1}%
  \BibitemOpen
  \bibfield  {author} {\bibinfo {author} {\bibfnamefont {B.}~\bibnamefont
  {Abbott}} \emph {et~al.} (\bibinfo {collaboration} {LIGO Scientific,
  Virgo}),\ }\href {\doibase 10.1103/PhysRevLett.116.061102} {\bibfield
  {journal} {\bibinfo  {journal} {Phys. Rev. Lett.}\ }\textbf {\bibinfo
  {volume} {116}},\ \bibinfo {pages} {061102} (\bibinfo {year}
  {2016}{\natexlab{b}})}\BibitemShut {NoStop}%
\bibitem [{\citenamefont {Abbott}\ \emph
  {et~al.}(2017{\natexlab{a}})\citenamefont {Abbott} \emph
  {et~al.}}]{ligo_PRL_2017}%
  \BibitemOpen
  \bibfield  {author} {\bibinfo {author} {\bibfnamefont {B.~P.}\ \bibnamefont
  {Abbott}} \emph {et~al.} (\bibinfo {collaboration} {LIGO Scientific,
  Virgo}),\ }\href {\doibase 10.1103/PhysRevLett.118.221101} {\bibfield
  {journal} {\bibinfo  {journal} {Phys. Rev. Lett.}\ }\textbf {\bibinfo
  {volume} {118}},\ \bibinfo {pages} {221101} (\bibinfo {year}
  {2017}{\natexlab{a}})},\ \bibinfo {note} {[Erratum: Phys.Rev.Lett. 121,
  129901 (2018)]}\BibitemShut {NoStop}%
\bibitem [{\citenamefont {Abbott}\ \emph
  {et~al.}(2017{\natexlab{b}})\citenamefont {Abbott} \emph
  {et~al.}}]{abbott_ApJL_2017}%
  \BibitemOpen
  \bibfield  {author} {\bibinfo {author} {\bibfnamefont {B.~P.}\ \bibnamefont
  {Abbott}} \emph {et~al.} (\bibinfo {collaboration} {LIGO Scientific,
  Virgo}),\ }\href {\doibase 10.3847/2041-8213/aa9f0c} {\bibfield  {journal}
  {\bibinfo  {journal} {Astrophys. J.}\ }\textbf {\bibinfo {volume} {851}},\
  \bibinfo {pages} {L35} (\bibinfo {year} {2017}{\natexlab{b}})}\BibitemShut
  {NoStop}%
\bibitem [{\citenamefont {Abbott}\ \emph
  {et~al.}(2017{\natexlab{c}})\citenamefont {Abbott} \emph
  {et~al.}}]{ligo_PRL_2017_1}%
  \BibitemOpen
  \bibfield  {author} {\bibinfo {author} {\bibfnamefont {B.}~\bibnamefont
  {Abbott}} \emph {et~al.} (\bibinfo {collaboration} {LIGO Scientific,
  Virgo}),\ }\href {\doibase 10.1103/PhysRevLett.119.141101} {\bibfield
  {journal} {\bibinfo  {journal} {Phys. Rev. Lett.}\ }\textbf {\bibinfo
  {volume} {119}},\ \bibinfo {pages} {141101} (\bibinfo {year}
  {2017}{\natexlab{c}})}\BibitemShut {NoStop}%
\bibitem [{\citenamefont {Abbott}\ \emph
  {et~al.}(2017{\natexlab{d}})\citenamefont {Abbott} \emph
  {et~al.}}]{ligo_PRL_2017_2}%
  \BibitemOpen
  \bibfield  {author} {\bibinfo {author} {\bibfnamefont {B.}~\bibnamefont
  {Abbott}} \emph {et~al.} (\bibinfo {collaboration} {LIGO Scientific,
  Virgo}),\ }\href {\doibase 10.1103/PhysRevLett.119.161101} {\bibfield
  {journal} {\bibinfo  {journal} {Phys. Rev. Lett.}\ }\textbf {\bibinfo
  {volume} {119}},\ \bibinfo {pages} {161101} (\bibinfo {year}
  {2017}{\natexlab{d}})}\BibitemShut {NoStop}%
\bibitem [{\citenamefont {Abbott}\ \emph {et~al.}(2019)\citenamefont {Abbott}
  \emph {et~al.}}]{ligo_PRX_2019}%
  \BibitemOpen
  \bibfield  {author} {\bibinfo {author} {\bibfnamefont {B.}~\bibnamefont
  {Abbott}} \emph {et~al.} (\bibinfo {collaboration} {LIGO Scientific,
  Virgo}),\ }\href {\doibase 10.1103/PhysRevX.9.031040} {\bibfield  {journal}
  {\bibinfo  {journal} {Phys. Rev. X}\ }\textbf {\bibinfo {volume} {9}},\
  \bibinfo {pages} {031040} (\bibinfo {year} {2019})}\BibitemShut {NoStop}%
\bibitem [{\citenamefont {Mann}\ \emph {et~al.}(2019)\citenamefont {Mann},
  \citenamefont {Richer}, \citenamefont {Heyl}, \citenamefont {Anderson},
  \citenamefont {Kalirai}, \citenamefont {Caiazzo}, \citenamefont {Möhle},
  \citenamefont {Knee},\ and\ \citenamefont {Baumgardt}}]{eht_ApJ_2019}%
  \BibitemOpen
  \bibfield  {author} {\bibinfo {author} {\bibfnamefont {C.~R.}\ \bibnamefont
  {Mann}}, \bibinfo {author} {\bibfnamefont {H.}~\bibnamefont {Richer}},
  \bibinfo {author} {\bibfnamefont {J.}~\bibnamefont {Heyl}}, \bibinfo {author}
  {\bibfnamefont {J.}~\bibnamefont {Anderson}}, \bibinfo {author}
  {\bibfnamefont {J.}~\bibnamefont {Kalirai}}, \bibinfo {author} {\bibfnamefont
  {I.}~\bibnamefont {Caiazzo}}, \bibinfo {author} {\bibfnamefont {S.~D.}\
  \bibnamefont {Möhle}}, \bibinfo {author} {\bibfnamefont {A.}~\bibnamefont
  {Knee}}, \ and\ \bibinfo {author} {\bibfnamefont {H.}~\bibnamefont
  {Baumgardt}},\ }\href {\doibase 10.3847/1538-4357/ab0e6d} {\bibfield
  {journal} {\bibinfo  {journal} {Astrophys. J.}\ }\textbf {\bibinfo {volume}
  {875}},\ \bibinfo {pages} {1} (\bibinfo {year} {2019})}\BibitemShut {NoStop}%
\bibitem [{\citenamefont {Martin}\ \emph {et~al.}(2019)\citenamefont {Martin},
  \citenamefont {Lubow}, \citenamefont {Pringle}, \citenamefont {Franchini},
  \citenamefont {Zhu}, \citenamefont {Lepp}, \citenamefont {Nealon},
  \citenamefont {Nixon},\ and\ \citenamefont {Vallet}}]{eht_ApJ_2019_1}%
  \BibitemOpen
  \bibfield  {author} {\bibinfo {author} {\bibfnamefont {R.~G.}\ \bibnamefont
  {Martin}}, \bibinfo {author} {\bibfnamefont {S.~H.}\ \bibnamefont {Lubow}},
  \bibinfo {author} {\bibfnamefont {J.~E.}\ \bibnamefont {Pringle}}, \bibinfo
  {author} {\bibfnamefont {A.}~\bibnamefont {Franchini}}, \bibinfo {author}
  {\bibfnamefont {Z.}~\bibnamefont {Zhu}}, \bibinfo {author} {\bibfnamefont
  {S.}~\bibnamefont {Lepp}}, \bibinfo {author} {\bibfnamefont {R.}~\bibnamefont
  {Nealon}}, \bibinfo {author} {\bibfnamefont {C.~J.}\ \bibnamefont {Nixon}}, \
  and\ \bibinfo {author} {\bibfnamefont {D.}~\bibnamefont {Vallet}},\ }\href
  {\doibase 10.3847/1538-4357/ab0bb7} {\bibfield  {journal} {\bibinfo
  {journal} {Astrophys. J.}\ }\textbf {\bibinfo {volume} {875}},\ \bibinfo
  {pages} {5} (\bibinfo {year} {2019})}\BibitemShut {NoStop}%
\bibitem [{\citenamefont {Zhu}\ \emph {et~al.}(2018)\citenamefont {Zhu},
  \citenamefont {Johnson},\ and\ \citenamefont {Narayan}}]{zhu_ApJ_2018}%
  \BibitemOpen
  \bibfield  {author} {\bibinfo {author} {\bibfnamefont {Z.}~\bibnamefont
  {Zhu}}, \bibinfo {author} {\bibfnamefont {M.~D.}\ \bibnamefont {Johnson}}, \
  and\ \bibinfo {author} {\bibfnamefont {R.}~\bibnamefont {Narayan}},\ }\href
  {\doibase 10.3847/1538-4357/aaef3d} {\bibfield  {journal} {\bibinfo
  {journal} {Astrophys. J.}\ }\textbf {\bibinfo {volume} {870}},\ \bibinfo
  {pages} {6} (\bibinfo {year} {2018})}\BibitemShut {NoStop}%
\bibitem [{\citenamefont {Eckart}\ \emph {et~al.}(2017)\citenamefont {Eckart},
  \citenamefont {Hüttemann}, \citenamefont {Kiefer}, \citenamefont {Britzen},
  \citenamefont {Zaja\v~cek}, \citenamefont {Lämmerzahl}, \citenamefont
  {Stöckler}, \citenamefont {Valencia-S}, \citenamefont {Karas},\ and\
  \citenamefont {García-Marín}}]{eckart_FP_2017}%
  \BibitemOpen
  \bibfield  {author} {\bibinfo {author} {\bibfnamefont {A.}~\bibnamefont
  {Eckart}}, \bibinfo {author} {\bibfnamefont {A.}~\bibnamefont {Hüttemann}},
  \bibinfo {author} {\bibfnamefont {C.}~\bibnamefont {Kiefer}}, \bibinfo
  {author} {\bibfnamefont {S.}~\bibnamefont {Britzen}}, \bibinfo {author}
  {\bibfnamefont {M.}~\bibnamefont {Zaja\v~cek}}, \bibinfo {author}
  {\bibfnamefont {C.}~\bibnamefont {Lämmerzahl}}, \bibinfo {author}
  {\bibfnamefont {M.}~\bibnamefont {Stöckler}}, \bibinfo {author}
  {\bibfnamefont {M.}~\bibnamefont {Valencia-S}}, \bibinfo {author}
  {\bibfnamefont {V.}~\bibnamefont {Karas}}, \ and\ \bibinfo {author}
  {\bibfnamefont {M.}~\bibnamefont {García-Marín}},\ }\href {\doibase
  10.1007/s10701-017-0079-2} {\bibfield  {journal} {\bibinfo  {journal} {Found.
  Phys.}\ }\textbf {\bibinfo {volume} {47}},\ \bibinfo {pages} {553} (\bibinfo
  {year} {2017})}\BibitemShut {NoStop}%
\bibitem [{\citenamefont {Abramowicz}\ \emph {et~al.}(2002)\citenamefont
  {Abramowicz}, \citenamefont {Kluzniak},\ and\ \citenamefont
  {Lasota}}]{abramowicz_AA_2002}%
  \BibitemOpen
  \bibfield  {author} {\bibinfo {author} {\bibfnamefont {M.~A.}\ \bibnamefont
  {Abramowicz}}, \bibinfo {author} {\bibfnamefont {W.}~\bibnamefont
  {Kluzniak}}, \ and\ \bibinfo {author} {\bibfnamefont {J.-P.}\ \bibnamefont
  {Lasota}},\ }\href {\doibase 10.1051/0004-6361:20021645} {\bibfield
  {journal} {\bibinfo  {journal} {Astron. Astrophys.}\ }\textbf {\bibinfo
  {volume} {396}},\ \bibinfo {pages} {L31} (\bibinfo {year}
  {2002})}\BibitemShut {NoStop}%
\bibitem [{\citenamefont {Unruh}\ and\ \citenamefont
  {Wald}(2017)}]{unruh_RPP_2017}%
  \BibitemOpen
  \bibfield  {author} {\bibinfo {author} {\bibfnamefont {W.~G.}\ \bibnamefont
  {Unruh}}\ and\ \bibinfo {author} {\bibfnamefont {R.~M.}\ \bibnamefont
  {Wald}},\ }\href {\doibase 10.1088/1361-6633/aa778e} {\bibfield  {journal}
  {\bibinfo  {journal} {Rept. Prog. Phys.}\ }\textbf {\bibinfo {volume} {80}},\
  \bibinfo {pages} {092002} (\bibinfo {year} {2017})}\BibitemShut {NoStop}%
\bibitem [{\citenamefont {Mazur}\ and\ \citenamefont
  {Mottola}(2001)}]{mazur_arxiv_2001}%
  \BibitemOpen
  \bibfield  {author} {\bibinfo {author} {\bibfnamefont {P.~O.}\ \bibnamefont
  {Mazur}}\ and\ \bibinfo {author} {\bibfnamefont {E.}~\bibnamefont
  {Mottola}},\ }\href@noop {} {\  (\bibinfo {year} {2001})},\ \Eprint
  {http://arxiv.org/abs/gr-qc/0109035} {arXiv:gr-qc/0109035} \BibitemShut
  {NoStop}%
\bibitem [{\citenamefont {Mazur}\ and\ \citenamefont
  {Mottola}(2004)}]{mazur_NAS_2004}%
  \BibitemOpen
  \bibfield  {author} {\bibinfo {author} {\bibfnamefont {P.~O.}\ \bibnamefont
  {Mazur}}\ and\ \bibinfo {author} {\bibfnamefont {E.}~\bibnamefont
  {Mottola}},\ }\href {\doibase 10.1073/pnas.0402717101} {\bibfield  {journal}
  {\bibinfo  {journal} {Proc. Natl. Acad. Sci. U.S.A.}\ }\textbf {\bibinfo
  {volume} {101}},\ \bibinfo {pages} {9545} (\bibinfo {year}
  {2004})}\BibitemShut {NoStop}%
\bibitem [{\citenamefont {Schunck}\ and\ \citenamefont
  {Mielke}(2003)}]{schunck_CQG_2003}%
  \BibitemOpen
  \bibfield  {author} {\bibinfo {author} {\bibfnamefont {F.~E.}\ \bibnamefont
  {Schunck}}\ and\ \bibinfo {author} {\bibfnamefont {E.~W.}\ \bibnamefont
  {Mielke}},\ }\href {\doibase 10.1088/0264-9381/20/20/201} {\bibfield
  {journal} {\bibinfo  {journal} {Class. Quantum Grav.}\ }\textbf {\bibinfo
  {volume} {20}},\ \bibinfo {pages} {R301} (\bibinfo {year}
  {2003})}\BibitemShut {NoStop}%
\bibitem [{\citenamefont {Morris}\ \emph {et~al.}(1988)\citenamefont {Morris},
  \citenamefont {Thorne},\ and\ \citenamefont {Yurtsever}}]{morris_PRL_1988}%
  \BibitemOpen
  \bibfield  {author} {\bibinfo {author} {\bibfnamefont {M.~S.}\ \bibnamefont
  {Morris}}, \bibinfo {author} {\bibfnamefont {K.~S.}\ \bibnamefont {Thorne}},
  \ and\ \bibinfo {author} {\bibfnamefont {U.}~\bibnamefont {Yurtsever}},\
  }\href {\doibase 10.1103/PhysRevLett.61.1446} {\bibfield  {journal} {\bibinfo
   {journal} {Phys. Rev. Lett.}\ }\textbf {\bibinfo {volume} {61}},\ \bibinfo
  {pages} {1446} (\bibinfo {year} {1988})}\BibitemShut {NoStop}%
\bibitem [{\citenamefont {Damour}\ and\ \citenamefont
  {Solodukhin}(2007)}]{damour_PRD_2007}%
  \BibitemOpen
  \bibfield  {author} {\bibinfo {author} {\bibfnamefont {T.}~\bibnamefont
  {Damour}}\ and\ \bibinfo {author} {\bibfnamefont {S.~N.}\ \bibnamefont
  {Solodukhin}},\ }\href {\doibase 10.1103/PhysRevD.76.024016} {\bibfield
  {journal} {\bibinfo  {journal} {Phys. Rev. D}\ }\textbf {\bibinfo {volume}
  {76}},\ \bibinfo {pages} {024016} (\bibinfo {year} {2007})}\BibitemShut
  {NoStop}%
\bibitem [{\citenamefont {Cardoso}\ \emph
  {et~al.}(2016{\natexlab{a}})\citenamefont {Cardoso}, \citenamefont
  {Franzin},\ and\ \citenamefont {Pani}}]{cardoso_PRL_2016}%
  \BibitemOpen
  \bibfield  {author} {\bibinfo {author} {\bibfnamefont {V.}~\bibnamefont
  {Cardoso}}, \bibinfo {author} {\bibfnamefont {E.}~\bibnamefont {Franzin}}, \
  and\ \bibinfo {author} {\bibfnamefont {P.}~\bibnamefont {Pani}},\ }\href
  {\doibase 10.1103/PhysRevLett.116.171101} {\bibfield  {journal} {\bibinfo
  {journal} {Phys. Rev. Lett.}\ }\textbf {\bibinfo {volume} {116}},\ \bibinfo
  {pages} {171101} (\bibinfo {year} {2016}{\natexlab{a}})},\ \bibinfo {note}
  {[Erratum: Phys.Rev.Lett. 117, 089902 (2016)]}\BibitemShut {NoStop}%
\bibitem [{\citenamefont {Mathur}(2005)}]{mathur_FP_2005}%
  \BibitemOpen
  \bibfield  {author} {\bibinfo {author} {\bibfnamefont {S.~D.}\ \bibnamefont
  {Mathur}},\ }\href {\doibase 10.1002/prop.200410203} {\bibfield  {journal}
  {\bibinfo  {journal} {Fortsch. Phys.}\ }\textbf {\bibinfo {volume} {53}},\
  \bibinfo {pages} {793} (\bibinfo {year} {2005})}\BibitemShut {NoStop}%
\bibitem [{\citenamefont {Barcelo}\ \emph {et~al.}(2011)\citenamefont
  {Barcelo}, \citenamefont {Garay},\ and\ \citenamefont
  {Jannes}}]{barcelo_FP_2011}%
  \BibitemOpen
  \bibfield  {author} {\bibinfo {author} {\bibfnamefont {C.}~\bibnamefont
  {Barcelo}}, \bibinfo {author} {\bibfnamefont {L.}~\bibnamefont {Garay}}, \
  and\ \bibinfo {author} {\bibfnamefont {G.}~\bibnamefont {Jannes}},\ }\href
  {\doibase 10.1007/s10701-011-9577-9} {\bibfield  {journal} {\bibinfo
  {journal} {Found. Phys.}\ }\textbf {\bibinfo {volume} {41}},\ \bibinfo
  {pages} {1532} (\bibinfo {year} {2011})}\BibitemShut {NoStop}%
\bibitem [{\citenamefont {Barcelo}\ \emph {et~al.}(2015)\citenamefont
  {Barcelo}, \citenamefont {Carballo-Rubio}, \citenamefont {Garay},\ and\
  \citenamefont {Jannes}}]{barcelo_CQG_2015}%
  \BibitemOpen
  \bibfield  {author} {\bibinfo {author} {\bibfnamefont {C.}~\bibnamefont
  {Barcelo}}, \bibinfo {author} {\bibfnamefont {R.}~\bibnamefont
  {Carballo-Rubio}}, \bibinfo {author} {\bibfnamefont {L.~J.}\ \bibnamefont
  {Garay}}, \ and\ \bibinfo {author} {\bibfnamefont {G.}~\bibnamefont
  {Jannes}},\ }\href {\doibase 10.1088/0264-9381/32/3/035012} {\bibfield
  {journal} {\bibinfo  {journal} {Class. Quantum Grav.}\ }\textbf {\bibinfo
  {volume} {32}},\ \bibinfo {pages} {035012} (\bibinfo {year}
  {2015})}\BibitemShut {NoStop}%
\bibitem [{\citenamefont {Holdom}\ and\ \citenamefont
  {Ren}(2017)}]{holdom_PRD_2017}%
  \BibitemOpen
  \bibfield  {author} {\bibinfo {author} {\bibfnamefont {B.}~\bibnamefont
  {Holdom}}\ and\ \bibinfo {author} {\bibfnamefont {J.}~\bibnamefont {Ren}},\
  }\href {\doibase 10.1103/PhysRevD.95.084034} {\bibfield  {journal} {\bibinfo
  {journal} {Phys. Rev. D}\ }\textbf {\bibinfo {volume} {95}},\ \bibinfo
  {pages} {084034} (\bibinfo {year} {2017})}\BibitemShut {NoStop}%
\bibitem [{\citenamefont {Konoplya}\ \emph
  {et~al.}(2019{\natexlab{a}})\citenamefont {Konoplya}, \citenamefont {Posada},
  \citenamefont {Stuchl\'{\i}k},\ and\ \citenamefont
  {Zhidenko}}]{konoplya_PRD_2019_1}%
  \BibitemOpen
  \bibfield  {author} {\bibinfo {author} {\bibfnamefont {R.~A.}\ \bibnamefont
  {Konoplya}}, \bibinfo {author} {\bibfnamefont {C.}~\bibnamefont {Posada}},
  \bibinfo {author} {\bibfnamefont {Z.}~\bibnamefont {Stuchl\'{\i}k}}, \ and\
  \bibinfo {author} {\bibfnamefont {A.}~\bibnamefont {Zhidenko}},\ }\href
  {\doibase 10.1103/PhysRevD.100.044027} {\bibfield  {journal} {\bibinfo
  {journal} {Phys. Rev. D}\ }\textbf {\bibinfo {volume} {100}},\ \bibinfo
  {pages} {044027} (\bibinfo {year} {2019}{\natexlab{a}})}\BibitemShut
  {NoStop}%
\bibitem [{\citenamefont {Posada}\ and\ \citenamefont
  {Chirenti}(2019)}]{posada_CQG_2019}%
  \BibitemOpen
  \bibfield  {author} {\bibinfo {author} {\bibfnamefont {C.}~\bibnamefont
  {Posada}}\ and\ \bibinfo {author} {\bibfnamefont {C.}~\bibnamefont
  {Chirenti}},\ }\href {\doibase 10.1088/1361-6382/ab0526} {\bibfield
  {journal} {\bibinfo  {journal} {Class. Quantum Grav.}\ }\textbf {\bibinfo
  {volume} {36}},\ \bibinfo {pages} {065004} (\bibinfo {year}
  {2019})}\BibitemShut {NoStop}%
\bibitem [{\citenamefont {Penrose}(2002)}]{penrose_GERG_2002}%
  \BibitemOpen
  \bibfield  {author} {\bibinfo {author} {\bibfnamefont {R.}~\bibnamefont
  {Penrose}},\ }\href {\doibase 10.1023/a:1016578408204} {\bibfield  {journal}
  {\bibinfo  {journal} {Gen. Rel. Grav.}\ }\textbf {\bibinfo {volume} {34}},\
  \bibinfo {pages} {1141} (\bibinfo {year} {2002})}\BibitemShut {NoStop}%
\bibitem [{\citenamefont {Liu}\ \emph {et~al.}(2014)\citenamefont {Liu},
  \citenamefont {Malafarina}, \citenamefont {Modesto},\ and\ \citenamefont
  {Bambi}}]{liu_PRD_2014}%
  \BibitemOpen
  \bibfield  {author} {\bibinfo {author} {\bibfnamefont {Y.}~\bibnamefont
  {Liu}}, \bibinfo {author} {\bibfnamefont {D.}~\bibnamefont {Malafarina}},
  \bibinfo {author} {\bibfnamefont {L.}~\bibnamefont {Modesto}}, \ and\
  \bibinfo {author} {\bibfnamefont {C.}~\bibnamefont {Bambi}},\ }\href
  {\doibase 10.1103/PhysRevD.90.044040} {\bibfield  {journal} {\bibinfo
  {journal} {Phys. Rev. D}\ }\textbf {\bibinfo {volume} {90}},\ \bibinfo
  {pages} {044040} (\bibinfo {year} {2014})}\BibitemShut {NoStop}%
\bibitem [{\citenamefont {Kiefer}\ and\ \citenamefont
  {Schmitz}(2019)}]{kiefer_PRD_2019}%
  \BibitemOpen
  \bibfield  {author} {\bibinfo {author} {\bibfnamefont {C.}~\bibnamefont
  {Kiefer}}\ and\ \bibinfo {author} {\bibfnamefont {T.}~\bibnamefont
  {Schmitz}},\ }\href {\doibase 10.1103/PhysRevD.99.126010} {\bibfield
  {journal} {\bibinfo  {journal} {Phys. Rev. D}\ }\textbf {\bibinfo {volume}
  {99}},\ \bibinfo {pages} {126010} (\bibinfo {year} {2019})}\BibitemShut
  {NoStop}%
\bibitem [{\citenamefont {Bojowald}\ \emph {et~al.}(2005)\citenamefont
  {Bojowald}, \citenamefont {Goswami}, \citenamefont {Maartens},\ and\
  \citenamefont {Singh}}]{bojowald_PRL_2005}%
  \BibitemOpen
  \bibfield  {author} {\bibinfo {author} {\bibfnamefont {M.}~\bibnamefont
  {Bojowald}}, \bibinfo {author} {\bibfnamefont {R.}~\bibnamefont {Goswami}},
  \bibinfo {author} {\bibfnamefont {R.}~\bibnamefont {Maartens}}, \ and\
  \bibinfo {author} {\bibfnamefont {P.}~\bibnamefont {Singh}},\ }\href
  {\doibase 10.1103/PhysRevLett.95.091302} {\bibfield  {journal} {\bibinfo
  {journal} {Phys. Rev. Lett.}\ }\textbf {\bibinfo {volume} {95}},\ \bibinfo
  {pages} {091302} (\bibinfo {year} {2005})}\BibitemShut {NoStop}%
\bibitem [{\citenamefont {Harada}\ \emph {et~al.}(2001)\citenamefont {Harada},
  \citenamefont {Iguchi}, \citenamefont {Nakao}, \citenamefont {Singh},
  \citenamefont {Tanaka},\ and\ \citenamefont {Vaz}}]{harada_PRD_2001}%
  \BibitemOpen
  \bibfield  {author} {\bibinfo {author} {\bibfnamefont {T.}~\bibnamefont
  {Harada}}, \bibinfo {author} {\bibfnamefont {H.}~\bibnamefont {Iguchi}},
  \bibinfo {author} {\bibfnamefont {K.-i.}\ \bibnamefont {Nakao}}, \bibinfo
  {author} {\bibfnamefont {T.~P.}\ \bibnamefont {Singh}}, \bibinfo {author}
  {\bibfnamefont {T.}~\bibnamefont {Tanaka}}, \ and\ \bibinfo {author}
  {\bibfnamefont {C.}~\bibnamefont {Vaz}},\ }\href {\doibase
  10.1103/PhysRevD.64.041501} {\bibfield  {journal} {\bibinfo  {journal} {Phys.
  Rev. D}\ }\textbf {\bibinfo {volume} {64}},\ \bibinfo {pages} {041501(R)}
  (\bibinfo {year} {2001})}\BibitemShut {NoStop}%
\bibitem [{\citenamefont {Choptuik}(1993)}]{choptuik_PRL_1993}%
  \BibitemOpen
  \bibfield  {author} {\bibinfo {author} {\bibfnamefont {M.~W.}\ \bibnamefont
  {Choptuik}},\ }\href {\doibase 10.1103/PhysRevLett.70.9} {\bibfield
  {journal} {\bibinfo  {journal} {Phys. Rev. Lett.}\ }\textbf {\bibinfo
  {volume} {70}},\ \bibinfo {pages} {9} (\bibinfo {year} {1993})}\BibitemShut
  {NoStop}%
\bibitem [{\citenamefont {Joshi}\ and\ \citenamefont
  {Dwivedi}(1993)}]{joshi_PRD_1993}%
  \BibitemOpen
  \bibfield  {author} {\bibinfo {author} {\bibfnamefont {P.~S.}\ \bibnamefont
  {Joshi}}\ and\ \bibinfo {author} {\bibfnamefont {I.~H.}\ \bibnamefont
  {Dwivedi}},\ }\href {\doibase 10.1103/physrevd.47.5357} {\bibfield  {journal}
  {\bibinfo  {journal} {Phys. Rev. D}\ }\textbf {\bibinfo {volume} {47}},\
  \bibinfo {pages} {5357} (\bibinfo {year} {1993})}\BibitemShut {NoStop}%
\bibitem [{\citenamefont {Waugh}\ and\ \citenamefont
  {Lake}(1988)}]{waugh_PRD_1988}%
  \BibitemOpen
  \bibfield  {author} {\bibinfo {author} {\bibfnamefont {B.}~\bibnamefont
  {Waugh}}\ and\ \bibinfo {author} {\bibfnamefont {K.}~\bibnamefont {Lake}},\
  }\href {\doibase 10.1103/physrevd.38.1315} {\bibfield  {journal} {\bibinfo
  {journal} {Phys. Rev. D}\ }\textbf {\bibinfo {volume} {38}},\ \bibinfo
  {pages} {1315} (\bibinfo {year} {1988})}\BibitemShut {NoStop}%
\bibitem [{\citenamefont {Christodoulou}(1984)}]{christodoulou_CMP_1984}%
  \BibitemOpen
  \bibfield  {author} {\bibinfo {author} {\bibfnamefont {D.}~\bibnamefont
  {Christodoulou}},\ }\href {\doibase 10.1007/bf01223743} {\bibfield  {journal}
  {\bibinfo  {journal} {Commun. Math. Phys.}\ }\textbf {\bibinfo {volume}
  {93}},\ \bibinfo {pages} {171} (\bibinfo {year} {1984})}\BibitemShut
  {NoStop}%
\bibitem [{\citenamefont {Eardley}\ and\ \citenamefont
  {Smarr}(1979)}]{eardley_PRD_1979}%
  \BibitemOpen
  \bibfield  {author} {\bibinfo {author} {\bibfnamefont {D.~M.}\ \bibnamefont
  {Eardley}}\ and\ \bibinfo {author} {\bibfnamefont {L.}~\bibnamefont
  {Smarr}},\ }\href {\doibase 10.1103/physrevd.19.2239} {\bibfield  {journal}
  {\bibinfo  {journal} {Phys. Rev. D}\ }\textbf {\bibinfo {volume} {19}},\
  \bibinfo {pages} {2239} (\bibinfo {year} {1979})}\BibitemShut {NoStop}%
\bibitem [{\citenamefont {Ori}\ and\ \citenamefont
  {Piran}(1987)}]{ori_PRL_1987}%
  \BibitemOpen
  \bibfield  {author} {\bibinfo {author} {\bibfnamefont {A.}~\bibnamefont
  {Ori}}\ and\ \bibinfo {author} {\bibfnamefont {T.}~\bibnamefont {Piran}},\
  }\href {\doibase 10.1103/physrevlett.59.2137} {\bibfield  {journal} {\bibinfo
   {journal} {Phys. Rev. Lett.}\ }\textbf {\bibinfo {volume} {59}},\ \bibinfo
  {pages} {2137} (\bibinfo {year} {1987})}\BibitemShut {NoStop}%
\bibitem [{\citenamefont {Giamb{\`{o}}}\ \emph {et~al.}(2004)\citenamefont
  {Giamb{\`{o}}}, \citenamefont {Giannoni}, \citenamefont {Magli},\ and\
  \citenamefont {Piccione}}]{giambo_GERG_2004}%
  \BibitemOpen
  \bibfield  {author} {\bibinfo {author} {\bibfnamefont {R.}~\bibnamefont
  {Giamb{\`{o}}}}, \bibinfo {author} {\bibfnamefont {F.}~\bibnamefont
  {Giannoni}}, \bibinfo {author} {\bibfnamefont {G.}~\bibnamefont {Magli}}, \
  and\ \bibinfo {author} {\bibfnamefont {P.}~\bibnamefont {Piccione}},\ }\href
  {\doibase 10.1023/b:gerg.0000022388.11306.e1} {\bibfield  {journal} {\bibinfo
   {journal} {Gen. Rel. Grav.}\ }\textbf {\bibinfo {volume} {36}},\ \bibinfo
  {pages} {1279} (\bibinfo {year} {2004})}\BibitemShut {NoStop}%
\bibitem [{\citenamefont {Shapiro}\ and\ \citenamefont
  {Teukolsky}(1991)}]{shapiro_PRL_1991}%
  \BibitemOpen
  \bibfield  {author} {\bibinfo {author} {\bibfnamefont {S.~L.}\ \bibnamefont
  {Shapiro}}\ and\ \bibinfo {author} {\bibfnamefont {S.~A.}\ \bibnamefont
  {Teukolsky}},\ }\href {\doibase 10.1103/physrevlett.66.994} {\bibfield
  {journal} {\bibinfo  {journal} {Phys. Rev. Lett.}\ }\textbf {\bibinfo
  {volume} {66}},\ \bibinfo {pages} {994} (\bibinfo {year} {1991})}\BibitemShut
  {NoStop}%
\bibitem [{\citenamefont {Lake}(1991)}]{lake_PRD_1991}%
  \BibitemOpen
  \bibfield  {author} {\bibinfo {author} {\bibfnamefont {K.}~\bibnamefont
  {Lake}},\ }\href {\doibase 10.1103/physrevd.43.1416} {\bibfield  {journal}
  {\bibinfo  {journal} {Phys. Rev. D}\ }\textbf {\bibinfo {volume} {43}},\
  \bibinfo {pages} {1416} (\bibinfo {year} {1991})}\BibitemShut {NoStop}%
\bibitem [{\citenamefont {Goswami}\ and\ \citenamefont
  {Joshi}(2007)}]{goswami_PRD_2007}%
  \BibitemOpen
  \bibfield  {author} {\bibinfo {author} {\bibfnamefont {R.}~\bibnamefont
  {Goswami}}\ and\ \bibinfo {author} {\bibfnamefont {P.~S.}\ \bibnamefont
  {Joshi}},\ }\href {\doibase 10.1103/PhysRevD.76.084026} {\bibfield  {journal}
  {\bibinfo  {journal} {Phys. Rev. D}\ }\textbf {\bibinfo {volume} {76}},\
  \bibinfo {pages} {084026} (\bibinfo {year} {2007})}\BibitemShut {NoStop}%
\bibitem [{\citenamefont {Joshi}\ \emph {et~al.}(2002)\citenamefont {Joshi},
  \citenamefont {Dadhich},\ and\ \citenamefont {Maartens}}]{joshi_PRD_2002}%
  \BibitemOpen
  \bibfield  {author} {\bibinfo {author} {\bibfnamefont {P.~S.}\ \bibnamefont
  {Joshi}}, \bibinfo {author} {\bibfnamefont {N.}~\bibnamefont {Dadhich}}, \
  and\ \bibinfo {author} {\bibfnamefont {R.}~\bibnamefont {Maartens}},\ }\href
  {\doibase 10.1103/PhysRevD.65.101501} {\bibfield  {journal} {\bibinfo
  {journal} {Phys. Rev. D}\ }\textbf {\bibinfo {volume} {65}},\ \bibinfo
  {pages} {101501(R)} (\bibinfo {year} {2002})}\BibitemShut {NoStop}%
\bibitem [{\citenamefont {Harada}\ \emph {et~al.}(1998)\citenamefont {Harada},
  \citenamefont {Iguchi},\ and\ \citenamefont {Nakao}}]{harada_PRD_1998}%
  \BibitemOpen
  \bibfield  {author} {\bibinfo {author} {\bibfnamefont {T.}~\bibnamefont
  {Harada}}, \bibinfo {author} {\bibfnamefont {H.}~\bibnamefont {Iguchi}}, \
  and\ \bibinfo {author} {\bibfnamefont {K.-i.}\ \bibnamefont {Nakao}},\ }\href
  {\doibase 10.1103/PhysRevD.58.041502} {\bibfield  {journal} {\bibinfo
  {journal} {Phys. Rev. D}\ }\textbf {\bibinfo {volume} {58}},\ \bibinfo
  {pages} {041502(R)} (\bibinfo {year} {1998})}\BibitemShut {NoStop}%
\bibitem [{\citenamefont {Banerjee}\ and\ \citenamefont
  {Chakrabarti}(2017)}]{banerjee_PRD_2017}%
  \BibitemOpen
  \bibfield  {author} {\bibinfo {author} {\bibfnamefont {N.}~\bibnamefont
  {Banerjee}}\ and\ \bibinfo {author} {\bibfnamefont {S.}~\bibnamefont
  {Chakrabarti}},\ }\href {\doibase 10.1103/PhysRevD.95.024015} {\bibfield
  {journal} {\bibinfo  {journal} {Phys. Rev. D}\ }\textbf {\bibinfo {volume}
  {95}},\ \bibinfo {pages} {024015} (\bibinfo {year} {2017})}\BibitemShut
  {NoStop}%
\bibitem [{\citenamefont {Bhattacharya}\ \emph {et~al.}(2020)\citenamefont
  {Bhattacharya}, \citenamefont {Dey}, \citenamefont {Mazumdar},\ and\
  \citenamefont {Sarkar}}]{bhattacharya_PRD_2020}%
  \BibitemOpen
  \bibfield  {author} {\bibinfo {author} {\bibfnamefont {K.}~\bibnamefont
  {Bhattacharya}}, \bibinfo {author} {\bibfnamefont {D.}~\bibnamefont {Dey}},
  \bibinfo {author} {\bibfnamefont {A.}~\bibnamefont {Mazumdar}}, \ and\
  \bibinfo {author} {\bibfnamefont {T.}~\bibnamefont {Sarkar}},\ }\href
  {\doibase 10.1103/PhysRevD.101.043005} {\bibfield  {journal} {\bibinfo
  {journal} {Phys. Rev. D}\ }\textbf {\bibinfo {volume} {101}},\ \bibinfo
  {pages} {043005} (\bibinfo {year} {2020})}\BibitemShut {NoStop}%
\bibitem [{\citenamefont {Hioki}\ and\ \citenamefont
  {Maeda}(2009)}]{hioki_PRD_2009}%
  \BibitemOpen
  \bibfield  {author} {\bibinfo {author} {\bibfnamefont {K.}~\bibnamefont
  {Hioki}}\ and\ \bibinfo {author} {\bibfnamefont {K.-i.}\ \bibnamefont
  {Maeda}},\ }\href {\doibase 10.1103/PhysRevD.80.024042} {\bibfield  {journal}
  {\bibinfo  {journal} {Phys. Rev. D}\ }\textbf {\bibinfo {volume} {80}},\
  \bibinfo {pages} {024042} (\bibinfo {year} {2009})}\BibitemShut {NoStop}%
\bibitem [{\citenamefont {Yang}\ and\ \citenamefont
  {Li}(2016)}]{yang_IJMPD_2016}%
  \BibitemOpen
  \bibfield  {author} {\bibinfo {author} {\bibfnamefont {L.}~\bibnamefont
  {Yang}}\ and\ \bibinfo {author} {\bibfnamefont {Z.}~\bibnamefont {Li}},\
  }\href {\doibase 10.1142/s0218271816500267} {\bibfield  {journal} {\bibinfo
  {journal} {Int. J. Mod. Phys. D}\ }\textbf {\bibinfo {volume} {25}},\
  \bibinfo {pages} {1650026} (\bibinfo {year} {2016})}\BibitemShut {NoStop}%
\bibitem [{\citenamefont {Takahashi}(2004)}]{takahashi_ApJ_2004}%
  \BibitemOpen
  \bibfield  {author} {\bibinfo {author} {\bibfnamefont {R.}~\bibnamefont
  {Takahashi}},\ }\href {\doibase 10.1086/422403} {\bibfield  {journal}
  {\bibinfo  {journal} {Astrophys. J.}\ }\textbf {\bibinfo {volume} {611}},\
  \bibinfo {pages} {996} (\bibinfo {year} {2004})}\BibitemShut {NoStop}%
\bibitem [{\citenamefont {Gyulchev}\ and\ \citenamefont
  {Yazadjiev}(2008)}]{gyulchev_PRD_2008}%
  \BibitemOpen
  \bibfield  {author} {\bibinfo {author} {\bibfnamefont {G.~N.}\ \bibnamefont
  {Gyulchev}}\ and\ \bibinfo {author} {\bibfnamefont {S.~S.}\ \bibnamefont
  {Yazadjiev}},\ }\href {\doibase 10.1103/PhysRevD.78.083004} {\bibfield
  {journal} {\bibinfo  {journal} {Phys. Rev. D}\ }\textbf {\bibinfo {volume}
  {78}},\ \bibinfo {pages} {083004} (\bibinfo {year} {2008})}\BibitemShut
  {NoStop}%
\bibitem [{\citenamefont {Virbhadra}\ and\ \citenamefont
  {Ellis}(2002)}]{virbhadra_PRD_2002}%
  \BibitemOpen
  \bibfield  {author} {\bibinfo {author} {\bibfnamefont {K.~S.}\ \bibnamefont
  {Virbhadra}}\ and\ \bibinfo {author} {\bibfnamefont {G.~F.~R.}\ \bibnamefont
  {Ellis}},\ }\href {\doibase 10.1103/PhysRevD.65.103004} {\bibfield  {journal}
  {\bibinfo  {journal} {Phys. Rev. D}\ }\textbf {\bibinfo {volume} {65}},\
  \bibinfo {pages} {103004} (\bibinfo {year} {2002})}\BibitemShut {NoStop}%
\bibitem [{\citenamefont {Virbhadra}\ \emph {et~al.}(1998)\citenamefont
  {Virbhadra}, \citenamefont {Narasimha},\ and\ \citenamefont
  {Chitre}}]{virbhadra_AA_1998}%
  \BibitemOpen
  \bibfield  {author} {\bibinfo {author} {\bibfnamefont {K.}~\bibnamefont
  {Virbhadra}}, \bibinfo {author} {\bibfnamefont {D.}~\bibnamefont
  {Narasimha}}, \ and\ \bibinfo {author} {\bibfnamefont {S.}~\bibnamefont
  {Chitre}},\ }\href@noop {} {\bibfield  {journal} {\bibinfo  {journal}
  {Astron. Astrophys.}\ }\textbf {\bibinfo {volume} {337}},\ \bibinfo {pages}
  {1} (\bibinfo {year} {1998})}\BibitemShut {NoStop}%
\bibitem [{\citenamefont {Virbhadra}\ and\ \citenamefont
  {Keeton}(2008)}]{virbhadra_PRD_2008}%
  \BibitemOpen
  \bibfield  {author} {\bibinfo {author} {\bibfnamefont {K.~S.}\ \bibnamefont
  {Virbhadra}}\ and\ \bibinfo {author} {\bibfnamefont {C.~R.}\ \bibnamefont
  {Keeton}},\ }\href {\doibase 10.1103/PhysRevD.77.124014} {\bibfield
  {journal} {\bibinfo  {journal} {Phys. Rev. D}\ }\textbf {\bibinfo {volume}
  {77}},\ \bibinfo {pages} {124014} (\bibinfo {year} {2008})}\BibitemShut
  {NoStop}%
\bibitem [{\citenamefont {Kov\'acs}\ and\ \citenamefont
  {Harko}(2010)}]{kovacs_PRD_2010}%
  \BibitemOpen
  \bibfield  {author} {\bibinfo {author} {\bibfnamefont {Z.}~\bibnamefont
  {Kov\'acs}}\ and\ \bibinfo {author} {\bibfnamefont {T.}~\bibnamefont
  {Harko}},\ }\href {\doibase 10.1103/PhysRevD.82.124047} {\bibfield  {journal}
  {\bibinfo  {journal} {Phys. Rev. D}\ }\textbf {\bibinfo {volume} {82}},\
  \bibinfo {pages} {124047} (\bibinfo {year} {2010})}\BibitemShut {NoStop}%
\bibitem [{\citenamefont {Blaschke}\ and\ \citenamefont
  {Stuchl\'{\i}k}(2016)}]{blaschke_PRD_2016}%
  \BibitemOpen
  \bibfield  {author} {\bibinfo {author} {\bibfnamefont {M.}~\bibnamefont
  {Blaschke}}\ and\ \bibinfo {author} {\bibfnamefont {Z.~c.~v.}\ \bibnamefont
  {Stuchl\'{\i}k}},\ }\href {\doibase 10.1103/PhysRevD.94.086006} {\bibfield
  {journal} {\bibinfo  {journal} {Phys. Rev. D}\ }\textbf {\bibinfo {volume}
  {94}},\ \bibinfo {pages} {086006} (\bibinfo {year} {2016})}\BibitemShut
  {NoStop}%
\bibitem [{\citenamefont {Stuchl{\'{\i}}k}\ and\ \citenamefont
  {Schee}(2010)}]{stuchlik_CQG_2010}%
  \BibitemOpen
  \bibfield  {author} {\bibinfo {author} {\bibfnamefont {Z.}~\bibnamefont
  {Stuchl{\'{\i}}k}}\ and\ \bibinfo {author} {\bibfnamefont {J.}~\bibnamefont
  {Schee}},\ }\href {\doibase 10.1088/0264-9381/27/21/215017} {\bibfield
  {journal} {\bibinfo  {journal} {Class. Quantum Grav.}\ }\textbf {\bibinfo
  {volume} {27}},\ \bibinfo {pages} {215017} (\bibinfo {year}
  {2010})}\BibitemShut {NoStop}%
\bibitem [{\citenamefont {Bambi}\ \emph {et~al.}(2009)\citenamefont {Bambi},
  \citenamefont {Freese}, \citenamefont {Harada}, \citenamefont {Takahashi},\
  and\ \citenamefont {Yoshida}}]{bambi_PRD_2009}%
  \BibitemOpen
  \bibfield  {author} {\bibinfo {author} {\bibfnamefont {C.}~\bibnamefont
  {Bambi}}, \bibinfo {author} {\bibfnamefont {K.}~\bibnamefont {Freese}},
  \bibinfo {author} {\bibfnamefont {T.}~\bibnamefont {Harada}}, \bibinfo
  {author} {\bibfnamefont {R.}~\bibnamefont {Takahashi}}, \ and\ \bibinfo
  {author} {\bibfnamefont {N.}~\bibnamefont {Yoshida}},\ }\href {\doibase
  10.1103/PhysRevD.80.104023} {\bibfield  {journal} {\bibinfo  {journal} {Phys.
  Rev. D}\ }\textbf {\bibinfo {volume} {80}},\ \bibinfo {pages} {104023}
  (\bibinfo {year} {2009})}\BibitemShut {NoStop}%
\bibitem [{\citenamefont {Joshi}\ \emph {et~al.}(2011)\citenamefont {Joshi},
  \citenamefont {Malafarina},\ and\ \citenamefont {Narayan}}]{joshi_CQG_2011}%
  \BibitemOpen
  \bibfield  {author} {\bibinfo {author} {\bibfnamefont {P.~S.}\ \bibnamefont
  {Joshi}}, \bibinfo {author} {\bibfnamefont {D.}~\bibnamefont {Malafarina}}, \
  and\ \bibinfo {author} {\bibfnamefont {R.}~\bibnamefont {Narayan}},\ }\href
  {\doibase 10.1088/0264-9381/28/23/235018} {\bibfield  {journal} {\bibinfo
  {journal} {Class. Quantum Grav.}\ }\textbf {\bibinfo {volume} {28}},\
  \bibinfo {pages} {235018} (\bibinfo {year} {2011})}\BibitemShut {NoStop}%
\bibitem [{\citenamefont {Pugliese}\ \emph {et~al.}(2011)\citenamefont
  {Pugliese}, \citenamefont {Quevedo},\ and\ \citenamefont
  {Ruffini}}]{pugliese_PRD_2011}%
  \BibitemOpen
  \bibfield  {author} {\bibinfo {author} {\bibfnamefont {D.}~\bibnamefont
  {Pugliese}}, \bibinfo {author} {\bibfnamefont {H.}~\bibnamefont {Quevedo}}, \
  and\ \bibinfo {author} {\bibfnamefont {R.}~\bibnamefont {Ruffini}},\ }\href
  {\doibase 10.1103/PhysRevD.84.044030} {\bibfield  {journal} {\bibinfo
  {journal} {Phys. Rev. D}\ }\textbf {\bibinfo {volume} {84}},\ \bibinfo
  {pages} {044030} (\bibinfo {year} {2011})}\BibitemShut {NoStop}%
\bibitem [{\citenamefont {Gyulchev}\ \emph {et~al.}(2019)\citenamefont
  {Gyulchev}, \citenamefont {Nedkova}, \citenamefont {Vetsov},\ and\
  \citenamefont {Yazadjiev}}]{gyulchev_PRD_2019}%
  \BibitemOpen
  \bibfield  {author} {\bibinfo {author} {\bibfnamefont {G.}~\bibnamefont
  {Gyulchev}}, \bibinfo {author} {\bibfnamefont {P.}~\bibnamefont {Nedkova}},
  \bibinfo {author} {\bibfnamefont {T.}~\bibnamefont {Vetsov}}, \ and\ \bibinfo
  {author} {\bibfnamefont {S.}~\bibnamefont {Yazadjiev}},\ }\href {\doibase
  10.1103/PhysRevD.100.024055} {\bibfield  {journal} {\bibinfo  {journal}
  {Phys. Rev. D}\ }\textbf {\bibinfo {volume} {100}},\ \bibinfo {pages}
  {024055} (\bibinfo {year} {2019})}\BibitemShut {NoStop}%
\bibitem [{\citenamefont {Chowdhury}\ \emph {et~al.}(2012)\citenamefont
  {Chowdhury}, \citenamefont {Patil}, \citenamefont {Malafarina},\ and\
  \citenamefont {Joshi}}]{chowdhury_PRD_2012}%
  \BibitemOpen
  \bibfield  {author} {\bibinfo {author} {\bibfnamefont {A.~N.}\ \bibnamefont
  {Chowdhury}}, \bibinfo {author} {\bibfnamefont {M.}~\bibnamefont {Patil}},
  \bibinfo {author} {\bibfnamefont {D.}~\bibnamefont {Malafarina}}, \ and\
  \bibinfo {author} {\bibfnamefont {P.~S.}\ \bibnamefont {Joshi}},\ }\href
  {\doibase 10.1103/PhysRevD.85.104031} {\bibfield  {journal} {\bibinfo
  {journal} {Phys. Rev. D}\ }\textbf {\bibinfo {volume} {85}},\ \bibinfo
  {pages} {104031} (\bibinfo {year} {2012})}\BibitemShut {NoStop}%
\bibitem [{\citenamefont {Sau}\ \emph {et~al.}(2020)\citenamefont {Sau},
  \citenamefont {Banerjee},\ and\ \citenamefont {SenGupta}}]{sau_PRD_2020}%
  \BibitemOpen
  \bibfield  {author} {\bibinfo {author} {\bibfnamefont {S.}~\bibnamefont
  {Sau}}, \bibinfo {author} {\bibfnamefont {I.}~\bibnamefont {Banerjee}}, \
  and\ \bibinfo {author} {\bibfnamefont {S.}~\bibnamefont {SenGupta}},\ }\href
  {\doibase 10.1103/PhysRevD.102.064027} {\bibfield  {journal} {\bibinfo
  {journal} {Phys. Rev. D}\ }\textbf {\bibinfo {volume} {102}},\ \bibinfo
  {pages} {064027} (\bibinfo {year} {2020})}\BibitemShut {NoStop}%
\bibitem [{\citenamefont {Kokkotas}\ and\ \citenamefont
  {Schmidt}(1999)}]{kokkotas_LRR_1999}%
  \BibitemOpen
  \bibfield  {author} {\bibinfo {author} {\bibfnamefont {K.~D.}\ \bibnamefont
  {Kokkotas}}\ and\ \bibinfo {author} {\bibfnamefont {B.~G.}\ \bibnamefont
  {Schmidt}},\ }\href {\doibase 10.12942/lrr-1999-2} {\bibfield  {journal}
  {\bibinfo  {journal} {Living Rev. in Rel.}\ }\textbf {\bibinfo {volume} {2}}
  (\bibinfo {year} {1999}),\ 10.12942/lrr-1999-2}\BibitemShut {NoStop}%
\bibitem [{\citenamefont {Berti}\ \emph {et~al.}(2009)\citenamefont {Berti},
  \citenamefont {Cardoso},\ and\ \citenamefont {Starinets}}]{berti_CQG_2009}%
  \BibitemOpen
  \bibfield  {author} {\bibinfo {author} {\bibfnamefont {E.}~\bibnamefont
  {Berti}}, \bibinfo {author} {\bibfnamefont {V.}~\bibnamefont {Cardoso}}, \
  and\ \bibinfo {author} {\bibfnamefont {A.~O.}\ \bibnamefont {Starinets}},\
  }\href {\doibase 10.1088/0264-9381/26/16/163001} {\bibfield  {journal}
  {\bibinfo  {journal} {Class. Quantum Grav.}\ }\textbf {\bibinfo {volume}
  {26}},\ \bibinfo {pages} {163001} (\bibinfo {year} {2009})}\BibitemShut
  {NoStop}%
\bibitem [{\citenamefont {DeBenedictis}\ \emph {et~al.}(2006)\citenamefont
  {DeBenedictis}, \citenamefont {Horvat}, \citenamefont {Iliji{\'{c}}},
  \citenamefont {Kloster},\ and\ \citenamefont
  {Viswanathan}}]{debenedictis_CQG_2006}%
  \BibitemOpen
  \bibfield  {author} {\bibinfo {author} {\bibfnamefont {A.}~\bibnamefont
  {DeBenedictis}}, \bibinfo {author} {\bibfnamefont {D.}~\bibnamefont
  {Horvat}}, \bibinfo {author} {\bibfnamefont {S.}~\bibnamefont
  {Iliji{\'{c}}}}, \bibinfo {author} {\bibfnamefont {S.}~\bibnamefont
  {Kloster}}, \ and\ \bibinfo {author} {\bibfnamefont {K.~S.}\ \bibnamefont
  {Viswanathan}},\ }\href {\doibase 10.1088/0264-9381/23/7/007} {\bibfield
  {journal} {\bibinfo  {journal} {Class. Quantum Grav.}\ }\textbf {\bibinfo
  {volume} {23}},\ \bibinfo {pages} {2303} (\bibinfo {year}
  {2006})}\BibitemShut {NoStop}%
\bibitem [{\citenamefont {Chirenti}\ and\ \citenamefont
  {Rezzolla}(2007)}]{chirenti_CQG_2007}%
  \BibitemOpen
  \bibfield  {author} {\bibinfo {author} {\bibfnamefont {C.~B. M.~H.}\
  \bibnamefont {Chirenti}}\ and\ \bibinfo {author} {\bibfnamefont
  {L.}~\bibnamefont {Rezzolla}},\ }\href {\doibase 10.1088/0264-9381/24/16/013}
  {\bibfield  {journal} {\bibinfo  {journal} {Class. Quantum Grav.}\ }\textbf
  {\bibinfo {volume} {24}},\ \bibinfo {pages} {4191} (\bibinfo {year}
  {2007})}\BibitemShut {NoStop}%
\bibitem [{\citenamefont {Pani}\ \emph {et~al.}(2009)\citenamefont {Pani},
  \citenamefont {Berti}, \citenamefont {Cardoso}, \citenamefont {Chen},\ and\
  \citenamefont {Norte}}]{pani_PRD_2009}%
  \BibitemOpen
  \bibfield  {author} {\bibinfo {author} {\bibfnamefont {P.}~\bibnamefont
  {Pani}}, \bibinfo {author} {\bibfnamefont {E.}~\bibnamefont {Berti}},
  \bibinfo {author} {\bibfnamefont {V.}~\bibnamefont {Cardoso}}, \bibinfo
  {author} {\bibfnamefont {Y.}~\bibnamefont {Chen}}, \ and\ \bibinfo {author}
  {\bibfnamefont {R.}~\bibnamefont {Norte}},\ }\href {\doibase
  10.1103/PhysRevD.80.124047} {\bibfield  {journal} {\bibinfo  {journal} {Phys.
  Rev. D}\ }\textbf {\bibinfo {volume} {80}},\ \bibinfo {pages} {124047}
  (\bibinfo {year} {2009})}\BibitemShut {NoStop}%
\bibitem [{\citenamefont {Ferrari}\ and\ \citenamefont
  {Kokkotas}(2000)}]{ferrari_PRD_2000}%
  \BibitemOpen
  \bibfield  {author} {\bibinfo {author} {\bibfnamefont {V.}~\bibnamefont
  {Ferrari}}\ and\ \bibinfo {author} {\bibfnamefont {K.~D.}\ \bibnamefont
  {Kokkotas}},\ }\href {\doibase 10.1103/PhysRevD.62.107504} {\bibfield
  {journal} {\bibinfo  {journal} {Phys. Rev. D}\ }\textbf {\bibinfo {volume}
  {62}},\ \bibinfo {pages} {107504} (\bibinfo {year} {2000})}\BibitemShut
  {NoStop}%
\bibitem [{\citenamefont {Chirenti}\ \emph {et~al.}(2012)\citenamefont
  {Chirenti}, \citenamefont {Saa},\ and\ \citenamefont
  {Sk\'akala}}]{chirenti_PRD_2012}%
  \BibitemOpen
  \bibfield  {author} {\bibinfo {author} {\bibfnamefont {C.}~\bibnamefont
  {Chirenti}}, \bibinfo {author} {\bibfnamefont {A.}~\bibnamefont {Saa}}, \
  and\ \bibinfo {author} {\bibfnamefont {J.}~\bibnamefont {Sk\'akala}},\ }\href
  {\doibase 10.1103/PhysRevD.86.124008} {\bibfield  {journal} {\bibinfo
  {journal} {Phys. Rev. D}\ }\textbf {\bibinfo {volume} {86}},\ \bibinfo
  {pages} {124008} (\bibinfo {year} {2012})}\BibitemShut {NoStop}%
\bibitem [{\citenamefont {Maggio}\ \emph {et~al.}(2020)\citenamefont {Maggio},
  \citenamefont {Buoninfante}, \citenamefont {Mazumdar},\ and\ \citenamefont
  {Pani}}]{maggio_arxiv_2020}%
  \BibitemOpen
  \bibfield  {author} {\bibinfo {author} {\bibfnamefont {E.}~\bibnamefont
  {Maggio}}, \bibinfo {author} {\bibfnamefont {L.}~\bibnamefont {Buoninfante}},
  \bibinfo {author} {\bibfnamefont {A.}~\bibnamefont {Mazumdar}}, \ and\
  \bibinfo {author} {\bibfnamefont {P.}~\bibnamefont {Pani}},\ }\href {\doibase
  10.1103/PhysRevD.102.064053} {\bibfield  {journal} {\bibinfo  {journal}
  {Phys. Rev. D}\ }\textbf {\bibinfo {volume} {102}},\ \bibinfo {pages}
  {064053} (\bibinfo {year} {2020})}\BibitemShut {NoStop}%
\bibitem [{\citenamefont {Cardoso}\ \emph
  {et~al.}(2016{\natexlab{b}})\citenamefont {Cardoso}, \citenamefont {Hopper},
  \citenamefont {Macedo}, \citenamefont {Palenzuela},\ and\ \citenamefont
  {Pani}}]{cardoso_PRD_2016}%
  \BibitemOpen
  \bibfield  {author} {\bibinfo {author} {\bibfnamefont {V.}~\bibnamefont
  {Cardoso}}, \bibinfo {author} {\bibfnamefont {S.}~\bibnamefont {Hopper}},
  \bibinfo {author} {\bibfnamefont {C.~F.~B.}\ \bibnamefont {Macedo}}, \bibinfo
  {author} {\bibfnamefont {C.}~\bibnamefont {Palenzuela}}, \ and\ \bibinfo
  {author} {\bibfnamefont {P.}~\bibnamefont {Pani}},\ }\href {\doibase
  10.1103/PhysRevD.94.084031} {\bibfield  {journal} {\bibinfo  {journal} {Phys.
  Rev. D}\ }\textbf {\bibinfo {volume} {94}},\ \bibinfo {pages} {084031}
  (\bibinfo {year} {2016}{\natexlab{b}})}\BibitemShut {NoStop}%
\bibitem [{\citenamefont {Cardoso}\ and\ \citenamefont
  {Pani}(2017)}]{cardoso_NA_2017}%
  \BibitemOpen
  \bibfield  {author} {\bibinfo {author} {\bibfnamefont {V.}~\bibnamefont
  {Cardoso}}\ and\ \bibinfo {author} {\bibfnamefont {P.}~\bibnamefont {Pani}},\
  }\href {\doibase 10.1038/s41550-017-0225-y} {\bibfield  {journal} {\bibinfo
  {journal} {Nature Astron.}\ }\textbf {\bibinfo {volume} {1}},\ \bibinfo
  {pages} {586} (\bibinfo {year} {2017})}\BibitemShut {NoStop}%
\bibitem [{\citenamefont {Mark}\ \emph {et~al.}(2017)\citenamefont {Mark},
  \citenamefont {Zimmerman}, \citenamefont {Du},\ and\ \citenamefont
  {Chen}}]{mark_PRD_2017}%
  \BibitemOpen
  \bibfield  {author} {\bibinfo {author} {\bibfnamefont {Z.}~\bibnamefont
  {Mark}}, \bibinfo {author} {\bibfnamefont {A.}~\bibnamefont {Zimmerman}},
  \bibinfo {author} {\bibfnamefont {S.~M.}\ \bibnamefont {Du}}, \ and\ \bibinfo
  {author} {\bibfnamefont {Y.}~\bibnamefont {Chen}},\ }\href {\doibase
  10.1103/PhysRevD.96.084002} {\bibfield  {journal} {\bibinfo  {journal} {Phys.
  Rev. D}\ }\textbf {\bibinfo {volume} {96}},\ \bibinfo {pages} {084002}
  (\bibinfo {year} {2017})}\BibitemShut {NoStop}%
\bibitem [{\citenamefont {Konoplya}\ \emph
  {et~al.}(2019{\natexlab{b}})\citenamefont {Konoplya}, \citenamefont
  {Stuchl\'{\i}k},\ and\ \citenamefont {Zhidenko}}]{konoplya_PRD_2019}%
  \BibitemOpen
  \bibfield  {author} {\bibinfo {author} {\bibfnamefont {R.~A.}\ \bibnamefont
  {Konoplya}}, \bibinfo {author} {\bibfnamefont {Z.}~\bibnamefont
  {Stuchl\'{\i}k}}, \ and\ \bibinfo {author} {\bibfnamefont {A.}~\bibnamefont
  {Zhidenko}},\ }\href {\doibase 10.1103/PhysRevD.99.024007} {\bibfield
  {journal} {\bibinfo  {journal} {Phys. Rev. D}\ }\textbf {\bibinfo {volume}
  {99}},\ \bibinfo {pages} {024007} (\bibinfo {year}
  {2019}{\natexlab{b}})}\BibitemShut {NoStop}%
\bibitem [{\citenamefont {Micchi}\ and\ \citenamefont
  {Chirenti}(2020)}]{chirenti_PRD_2020}%
  \BibitemOpen
  \bibfield  {author} {\bibinfo {author} {\bibfnamefont {L.~F.~L.}\
  \bibnamefont {Micchi}}\ and\ \bibinfo {author} {\bibfnamefont
  {C.}~\bibnamefont {Chirenti}},\ }\href {\doibase 10.1103/PhysRevD.101.084010}
  {\bibfield  {journal} {\bibinfo  {journal} {Phys. Rev. D}\ }\textbf {\bibinfo
  {volume} {101}},\ \bibinfo {pages} {084010} (\bibinfo {year}
  {2020})}\BibitemShut {NoStop}%
\bibitem [{\citenamefont {Churilova}\ and\ \citenamefont
  {Stuchl{\'{\i}}k}(2020)}]{churilova_CQG_2020}%
  \BibitemOpen
  \bibfield  {author} {\bibinfo {author} {\bibfnamefont {M.~S.}\ \bibnamefont
  {Churilova}}\ and\ \bibinfo {author} {\bibfnamefont {Z.}~\bibnamefont
  {Stuchl{\'{\i}}k}},\ }\href {\doibase 10.1088/1361-6382/ab7717} {\bibfield
  {journal} {\bibinfo  {journal} {Class. Quantum Grav.}\ }\textbf {\bibinfo
  {volume} {37}},\ \bibinfo {pages} {075014} (\bibinfo {year}
  {2020})}\BibitemShut {NoStop}%
\bibitem [{\citenamefont {Bronnikov}\ and\ \citenamefont
  {Konoplya}(2020)}]{bronnikov_PRD_2020}%
  \BibitemOpen
  \bibfield  {author} {\bibinfo {author} {\bibfnamefont {K.~A.}\ \bibnamefont
  {Bronnikov}}\ and\ \bibinfo {author} {\bibfnamefont {R.~A.}\ \bibnamefont
  {Konoplya}},\ }\href {\doibase 10.1103/PhysRevD.101.064004} {\bibfield
  {journal} {\bibinfo  {journal} {Phys. Rev. D}\ }\textbf {\bibinfo {volume}
  {101}},\ \bibinfo {pages} {064004} (\bibinfo {year} {2020})}\BibitemShut
  {NoStop}%
\bibitem [{\citenamefont {Dutta~Roy}\ \emph {et~al.}(2020)\citenamefont
  {Dutta~Roy}, \citenamefont {Aneesh},\ and\ \citenamefont
  {Kar}}]{roy_arxiv_2019}%
  \BibitemOpen
  \bibfield  {author} {\bibinfo {author} {\bibfnamefont {P.}~\bibnamefont
  {Dutta~Roy}}, \bibinfo {author} {\bibfnamefont {S.}~\bibnamefont {Aneesh}}, \
  and\ \bibinfo {author} {\bibfnamefont {S.}~\bibnamefont {Kar}},\ }\href
  {\doibase 10.1140/epjc/s10052-020-8409-5} {\bibfield  {journal} {\bibinfo
  {journal} {Eur. Phys. J. C}\ }\textbf {\bibinfo {volume} {80}},\ \bibinfo
  {pages} {850} (\bibinfo {year} {2020})}\BibitemShut {NoStop}%
\bibitem [{\citenamefont {Tsang}\ \emph {et~al.}(2018)\citenamefont {Tsang},
  \citenamefont {Rollier}, \citenamefont {Ghosh}, \citenamefont {Samajdar},
  \citenamefont {Agathos}, \citenamefont {Chatziioannou}, \citenamefont
  {Cardoso}, \citenamefont {Khanna},\ and\ \citenamefont {Van
  Den~Broeck}}]{tsang_PRD_2018}%
  \BibitemOpen
  \bibfield  {author} {\bibinfo {author} {\bibfnamefont {K.~W.}\ \bibnamefont
  {Tsang}}, \bibinfo {author} {\bibfnamefont {M.}~\bibnamefont {Rollier}},
  \bibinfo {author} {\bibfnamefont {A.}~\bibnamefont {Ghosh}}, \bibinfo
  {author} {\bibfnamefont {A.}~\bibnamefont {Samajdar}}, \bibinfo {author}
  {\bibfnamefont {M.}~\bibnamefont {Agathos}}, \bibinfo {author} {\bibfnamefont
  {K.}~\bibnamefont {Chatziioannou}}, \bibinfo {author} {\bibfnamefont
  {V.}~\bibnamefont {Cardoso}}, \bibinfo {author} {\bibfnamefont
  {G.}~\bibnamefont {Khanna}}, \ and\ \bibinfo {author} {\bibfnamefont
  {C.}~\bibnamefont {Van Den~Broeck}},\ }\href {\doibase
  10.1103/PhysRevD.98.024023} {\bibfield  {journal} {\bibinfo  {journal} {Phys.
  Rev. D}\ }\textbf {\bibinfo {volume} {98}},\ \bibinfo {pages} {024023}
  (\bibinfo {year} {2018})}\BibitemShut {NoStop}%
\bibitem [{\citenamefont {Janis}\ \emph {et~al.}(1968)\citenamefont {Janis},
  \citenamefont {Newman},\ and\ \citenamefont {Winicour}}]{janis_PRL_1968}%
  \BibitemOpen
  \bibfield  {author} {\bibinfo {author} {\bibfnamefont {A.~I.}\ \bibnamefont
  {Janis}}, \bibinfo {author} {\bibfnamefont {E.~T.}\ \bibnamefont {Newman}}, \
  and\ \bibinfo {author} {\bibfnamefont {J.}~\bibnamefont {Winicour}},\ }\href
  {\doibase 10.1103/physrevlett.20.878} {\bibfield  {journal} {\bibinfo
  {journal} {Phys. Rev. Lett.}\ }\textbf {\bibinfo {volume} {20}},\ \bibinfo
  {pages} {878} (\bibinfo {year} {1968})}\BibitemShut {NoStop}%
\bibitem [{\citenamefont {Wyman}(1981)}]{wyman_PRD_1981}%
  \BibitemOpen
  \bibfield  {author} {\bibinfo {author} {\bibfnamefont {M.}~\bibnamefont
  {Wyman}},\ }\href {\doibase 10.1103/physrevd.24.839} {\bibfield  {journal}
  {\bibinfo  {journal} {Phys. Rev. D}\ }\textbf {\bibinfo {volume} {24}},\
  \bibinfo {pages} {839} (\bibinfo {year} {1981})}\BibitemShut {NoStop}%
\bibitem [{\citenamefont {Virbhadra}(1997)}]{virbhadra_IJMPA_1997}%
  \BibitemOpen
  \bibfield  {author} {\bibinfo {author} {\bibfnamefont {K.~S.}\ \bibnamefont
  {Virbhadra}},\ }\href {\doibase 10.1142/s0217751x97002577} {\bibfield
  {journal} {\bibinfo  {journal} {Int. J. Mod. Phys. A}\ }\textbf {\bibinfo
  {volume} {12}},\ \bibinfo {pages} {4831} (\bibinfo {year}
  {1997})}\BibitemShut {NoStop}%
\bibitem [{\citenamefont {Jusufi}\ \emph {et~al.}(2019)\citenamefont {Jusufi},
  \citenamefont {Banerjee}, \citenamefont {Gyulchev},\ and\ \citenamefont
  {Amir}}]{jusufi_EPJC_2019}%
  \BibitemOpen
  \bibfield  {author} {\bibinfo {author} {\bibfnamefont {K.}~\bibnamefont
  {Jusufi}}, \bibinfo {author} {\bibfnamefont {A.}~\bibnamefont {Banerjee}},
  \bibinfo {author} {\bibfnamefont {G.}~\bibnamefont {Gyulchev}}, \ and\
  \bibinfo {author} {\bibfnamefont {M.}~\bibnamefont {Amir}},\ }\href {\doibase
  10.1140/epjc/s10052-019-6557-2} {\bibfield  {journal} {\bibinfo  {journal}
  {Eur. Phys. J. C}\ }\textbf {\bibinfo {volume} {79}},\ \bibinfo {pages} {28}
  (\bibinfo {year} {2019})}\BibitemShut {NoStop}%
\bibitem [{\citenamefont {Liu}\ \emph {et~al.}(2018)\citenamefont {Liu},
  \citenamefont {Zhou},\ and\ \citenamefont {Bambi}}]{liu_JCAP_2018}%
  \BibitemOpen
  \bibfield  {author} {\bibinfo {author} {\bibfnamefont {H.}~\bibnamefont
  {Liu}}, \bibinfo {author} {\bibfnamefont {M.}~\bibnamefont {Zhou}}, \ and\
  \bibinfo {author} {\bibfnamefont {C.}~\bibnamefont {Bambi}},\ }\href
  {\doibase 10.1088/1475-7516/2018/08/044} {\bibfield  {journal} {\bibinfo
  {journal} {JCAP}\ }\textbf {\bibinfo {volume} {2018}},\ \bibinfo {pages}
  {044} (\bibinfo {year} {2018})}\BibitemShut {NoStop}%
\bibitem [{\citenamefont {Sadhu}\ and\ \citenamefont
  {Suneeta}(2013)}]{varadrajan_IJMPD_2012}%
  \BibitemOpen
  \bibfield  {author} {\bibinfo {author} {\bibfnamefont {A.}~\bibnamefont
  {Sadhu}}\ and\ \bibinfo {author} {\bibfnamefont {V.}~\bibnamefont
  {Suneeta}},\ }\href {\doibase 10.1142/s0218271813500156} {\bibfield
  {journal} {\bibinfo  {journal} {Int. J. Mod. Phys. D}\ }\textbf {\bibinfo
  {volume} {22}},\ \bibinfo {pages} {1350015} (\bibinfo {year}
  {2013})}\BibitemShut {NoStop}%
\bibitem [{\citenamefont {Liao}\ \emph {et~al.}(2014)\citenamefont {Liao},
  \citenamefont {Chen}, \citenamefont {Huang},\ and\ \citenamefont
  {Wang}}]{liao_GERG_2014}%
  \BibitemOpen
  \bibfield  {author} {\bibinfo {author} {\bibfnamefont {P.}~\bibnamefont
  {Liao}}, \bibinfo {author} {\bibfnamefont {J.}~\bibnamefont {Chen}}, \bibinfo
  {author} {\bibfnamefont {H.}~\bibnamefont {Huang}}, \ and\ \bibinfo {author}
  {\bibfnamefont {Y.}~\bibnamefont {Wang}},\ }\href
  {http://dx.doi.org/10.1007/s10714-014-1752-9} {\bibfield  {journal} {\bibinfo
   {journal} {Gen. Rel. Grav.}\ }\textbf {\bibinfo {volume} {46}} (\bibinfo
  {year} {2014})}\BibitemShut {NoStop}%
\bibitem [{\citenamefont {Dey}\ \emph {et~al.}(2013)\citenamefont {Dey},
  \citenamefont {Roy},\ and\ \citenamefont {Sarkar}}]{dey_arxiv_2013}%
  \BibitemOpen
  \bibfield  {author} {\bibinfo {author} {\bibfnamefont {A.}~\bibnamefont
  {Dey}}, \bibinfo {author} {\bibfnamefont {P.}~\bibnamefont {Roy}}, \ and\
  \bibinfo {author} {\bibfnamefont {T.}~\bibnamefont {Sarkar}},\ }\href@noop {}
  {\  (\bibinfo {year} {2013})},\ \Eprint {http://arxiv.org/abs/1303.6824}
  {arXiv:1303.6824 [gr-qc]} \BibitemShut {NoStop}%
\bibitem [{\citenamefont {Chirenti}\ \emph {et~al.}(2013)\citenamefont
  {Chirenti}, \citenamefont {Saa},\ and\ \citenamefont
  {Sk\'{a}kala}}]{chirenti_PRD_2013}%
  \BibitemOpen
  \bibfield  {author} {\bibinfo {author} {\bibfnamefont {C.}~\bibnamefont
  {Chirenti}}, \bibinfo {author} {\bibfnamefont {A.}~\bibnamefont {Saa}}, \
  and\ \bibinfo {author} {\bibfnamefont {J.}~\bibnamefont {Sk\'{a}kala}},\
  }\href {\doibase 10.1103/PhysRevD.87.044034} {\bibfield  {journal} {\bibinfo
  {journal} {Phys. Rev. D}\ }\textbf {\bibinfo {volume} {87}},\ \bibinfo
  {pages} {044034} (\bibinfo {year} {2013})}\BibitemShut {NoStop}%
\bibitem [{\citenamefont {Virbhadra}\ \emph {et~al.}(1997)\citenamefont
  {Virbhadra}, \citenamefont {Jhingan},\ and\ \citenamefont
  {Joshi}}]{virbhadra_IJMPD_1997}%
  \BibitemOpen
  \bibfield  {author} {\bibinfo {author} {\bibfnamefont {K.~S.}\ \bibnamefont
  {Virbhadra}}, \bibinfo {author} {\bibfnamefont {S.}~\bibnamefont {Jhingan}},
  \ and\ \bibinfo {author} {\bibfnamefont {P.~S.}\ \bibnamefont {Joshi}},\
  }\href {\doibase 10.1142/s0218271897000200} {\bibfield  {journal} {\bibinfo
  {journal} {Int. J. Mod. Phys. D}\ }\textbf {\bibinfo {volume} {06}},\
  \bibinfo {pages} {357} (\bibinfo {year} {1997})}\BibitemShut {NoStop}%
\bibitem [{\citenamefont {Claudel}\ \emph {et~al.}(2001)\citenamefont
  {Claudel}, \citenamefont {Virbhadra},\ and\ \citenamefont
  {Ellis}}]{claudel_JMP_2001}%
  \BibitemOpen
  \bibfield  {author} {\bibinfo {author} {\bibfnamefont {C.-M.}\ \bibnamefont
  {Claudel}}, \bibinfo {author} {\bibfnamefont {K.~S.}\ \bibnamefont
  {Virbhadra}}, \ and\ \bibinfo {author} {\bibfnamefont {G.~F.~R.}\
  \bibnamefont {Ellis}},\ }\href {\doibase 10.1063/1.1308507} {\bibfield
  {journal} {\bibinfo  {journal} {J. Math. Phys.}\ }\textbf {\bibinfo {volume}
  {42}},\ \bibinfo {pages} {818} (\bibinfo {year} {2001})}\BibitemShut
  {NoStop}%
\bibitem [{\citenamefont {Regge}\ and\ \citenamefont
  {Wheeler}(1957)}]{regge_PRD_1957}%
  \BibitemOpen
  \bibfield  {author} {\bibinfo {author} {\bibfnamefont {T.}~\bibnamefont
  {Regge}}\ and\ \bibinfo {author} {\bibfnamefont {J.~A.}\ \bibnamefont
  {Wheeler}},\ }\href {\doibase 10.1103/physrev.108.1063} {\bibfield  {journal}
  {\bibinfo  {journal} {Phys. Rev.}\ }\textbf {\bibinfo {volume} {108}},\
  \bibinfo {pages} {1063} (\bibinfo {year} {1957})}\BibitemShut {NoStop}%
\bibitem [{\citenamefont {Zerilli}(1970)}]{zerilli_PRL_1970}%
  \BibitemOpen
  \bibfield  {author} {\bibinfo {author} {\bibfnamefont {F.~J.}\ \bibnamefont
  {Zerilli}},\ }\href {\doibase 10.1103/physrevlett.24.737} {\bibfield
  {journal} {\bibinfo  {journal} {Phys. Rev. Lett.}\ }\textbf {\bibinfo
  {volume} {24}},\ \bibinfo {pages} {737} (\bibinfo {year} {1970})}\BibitemShut
  {NoStop}%
\bibitem [{\citenamefont {Rezzolla}(2003)}]{rezzolla_review_2003}%
  \BibitemOpen
  \bibfield  {author} {\bibinfo {author} {\bibfnamefont {L.}~\bibnamefont
  {Rezzolla}},\ }\href@noop {} {\bibfield  {journal} {\bibinfo  {journal} {ICTP
  Lect. Notes Ser.}\ }\textbf {\bibinfo {volume} {14}},\ \bibinfo {pages} {255}
  (\bibinfo {year} {2003})}\BibitemShut {NoStop}%
\bibitem [{\citenamefont {Kobayashi}\ \emph {et~al.}(2012)\citenamefont
  {Kobayashi}, \citenamefont {Motohashi},\ and\ \citenamefont
  {Suyama}}]{kobayashi_PRD_2012}%
  \BibitemOpen
  \bibfield  {author} {\bibinfo {author} {\bibfnamefont {T.}~\bibnamefont
  {Kobayashi}}, \bibinfo {author} {\bibfnamefont {H.}~\bibnamefont
  {Motohashi}}, \ and\ \bibinfo {author} {\bibfnamefont {T.}~\bibnamefont
  {Suyama}},\ }\href {\doibase 10.1103/PhysRevD.85.084025} {\bibfield
  {journal} {\bibinfo  {journal} {Phys. Rev. D}\ }\textbf {\bibinfo {volume}
  {85}},\ \bibinfo {pages} {084025} (\bibinfo {year} {2012})}\BibitemShut
  {NoStop}%
\bibitem [{\citenamefont {Gundlach}\ \emph {et~al.}(1994)\citenamefont
  {Gundlach}, \citenamefont {Price},\ and\ \citenamefont
  {Pullin}}]{gundlach_PRD_1994}%
  \BibitemOpen
  \bibfield  {author} {\bibinfo {author} {\bibfnamefont {C.}~\bibnamefont
  {Gundlach}}, \bibinfo {author} {\bibfnamefont {R.~H.}\ \bibnamefont {Price}},
  \ and\ \bibinfo {author} {\bibfnamefont {J.}~\bibnamefont {Pullin}},\ }\href
  {\doibase 10.1103/physrevd.49.883} {\bibfield  {journal} {\bibinfo  {journal}
  {Phys. Rev. D}\ }\textbf {\bibinfo {volume} {49}},\ \bibinfo {pages} {883}
  (\bibinfo {year} {1994})}\BibitemShut {NoStop}%
\bibitem [{\citenamefont {Wald}(1980)}]{wald_JMP_1980}%
  \BibitemOpen
  \bibfield  {author} {\bibinfo {author} {\bibfnamefont {R.~M.}\ \bibnamefont
  {Wald}},\ }\href {\doibase 10.1063/1.524403} {\bibfield  {journal} {\bibinfo
  {journal} {J. Math. Phys.}\ }\textbf {\bibinfo {volume} {21}},\ \bibinfo
  {pages} {2802} (\bibinfo {year} {1980})}\BibitemShut {NoStop}%
\bibitem [{\citenamefont {Horowitz}\ and\ \citenamefont
  {Marolf}(1995)}]{horowitz_PRD_1995}%
  \BibitemOpen
  \bibfield  {author} {\bibinfo {author} {\bibfnamefont {G.~T.}\ \bibnamefont
  {Horowitz}}\ and\ \bibinfo {author} {\bibfnamefont {D.}~\bibnamefont
  {Marolf}},\ }\href {\doibase 10.1103/PhysRevD.52.5670} {\bibfield  {journal}
  {\bibinfo  {journal} {Phys. Rev. D}\ }\textbf {\bibinfo {volume} {52}},\
  \bibinfo {pages} {5670} (\bibinfo {year} {1995})}\BibitemShut {NoStop}%
\bibitem [{\citenamefont {Ishibashi}\ and\ \citenamefont
  {Hosoya}(1999)}]{ishibashi_PRD_1999}%
  \BibitemOpen
  \bibfield  {author} {\bibinfo {author} {\bibfnamefont {A.}~\bibnamefont
  {Ishibashi}}\ and\ \bibinfo {author} {\bibfnamefont {A.}~\bibnamefont
  {Hosoya}},\ }\href {\doibase 10.1103/PhysRevD.60.104028} {\bibfield
  {journal} {\bibinfo  {journal} {Phys. Rev. D}\ }\textbf {\bibinfo {volume}
  {60}},\ \bibinfo {pages} {104028} (\bibinfo {year} {1999})}\BibitemShut
  {NoStop}%
\bibitem [{\citenamefont {Ishibashi}\ and\ \citenamefont
  {Wald}(2003)}]{ishibashi_CQG_2003}%
  \BibitemOpen
  \bibfield  {author} {\bibinfo {author} {\bibfnamefont {A.}~\bibnamefont
  {Ishibashi}}\ and\ \bibinfo {author} {\bibfnamefont {R.~M.}\ \bibnamefont
  {Wald}},\ }\href {\doibase 10.1088/0264-9381/20/16/318} {\bibfield  {journal}
  {\bibinfo  {journal} {Class. Quantum Grav.}\ }\textbf {\bibinfo {volume}
  {20}},\ \bibinfo {pages} {3815} (\bibinfo {year} {2003})}\BibitemShut
  {NoStop}%
\bibitem [{\citenamefont {Helliwell}\ \emph {et~al.}(2003)\citenamefont
  {Helliwell}, \citenamefont {Konkowski},\ and\ \citenamefont
  {Arndt}}]{helliwell_GERG_2003}%
  \BibitemOpen
  \bibfield  {author} {\bibinfo {author} {\bibfnamefont {T.}~\bibnamefont
  {Helliwell}}, \bibinfo {author} {\bibfnamefont {D.}~\bibnamefont
  {Konkowski}}, \ and\ \bibinfo {author} {\bibfnamefont {V.}~\bibnamefont
  {Arndt}},\ }\href {\doibase 10.1023/A:1021307012363} {\bibfield  {journal}
  {\bibinfo  {journal} {Gen. Rel. Grav.}\ }\textbf {\bibinfo {volume} {35}},\
  \bibinfo {pages} {79} (\bibinfo {year} {2003})}\BibitemShut {NoStop}%
\bibitem [{\citenamefont {Ishibashi}\ and\ \citenamefont
  {Wald}(2004)}]{ishibashi_CQG_2004}%
  \BibitemOpen
  \bibfield  {author} {\bibinfo {author} {\bibfnamefont {A.}~\bibnamefont
  {Ishibashi}}\ and\ \bibinfo {author} {\bibfnamefont {R.~M.}\ \bibnamefont
  {Wald}},\ }\href {\doibase 10.1088/0264-9381/21/12/012} {\bibfield  {journal}
  {\bibinfo  {journal} {Class. Quantum Grav.}\ }\textbf {\bibinfo {volume}
  {21}},\ \bibinfo {pages} {2981} (\bibinfo {year} {2004})}\BibitemShut
  {NoStop}%
\bibitem [{\citenamefont {Gibbons}\ \emph {et~al.}(2005)\citenamefont
  {Gibbons}, \citenamefont {Hartnoll},\ and\ \citenamefont
  {Ishibashi}}]{gibbons_PTP_2005}%
  \BibitemOpen
  \bibfield  {author} {\bibinfo {author} {\bibfnamefont {G.~W.}\ \bibnamefont
  {Gibbons}}, \bibinfo {author} {\bibfnamefont {S.~A.}\ \bibnamefont
  {Hartnoll}}, \ and\ \bibinfo {author} {\bibfnamefont {A.}~\bibnamefont
  {Ishibashi}},\ }\href {\doibase 10.1143/PTP.113.963} {\bibfield  {journal}
  {\bibinfo  {journal} {Prog. Theor. Phys.}\ }\textbf {\bibinfo {volume}
  {113}},\ \bibinfo {pages} {963} (\bibinfo {year} {2005})}\BibitemShut
  {NoStop}%
\bibitem [{\citenamefont {Cardoso}\ and\ \citenamefont
  {Cavaglia}(2006)}]{cardoso_PRD_2006}%
  \BibitemOpen
  \bibfield  {author} {\bibinfo {author} {\bibfnamefont {V.}~\bibnamefont
  {Cardoso}}\ and\ \bibinfo {author} {\bibfnamefont {M.}~\bibnamefont
  {Cavaglia}},\ }\href {\doibase 10.1103/PhysRevD.74.024027} {\bibfield
  {journal} {\bibinfo  {journal} {Phys. Rev. D}\ }\textbf {\bibinfo {volume}
  {74}},\ \bibinfo {pages} {024027} (\bibinfo {year} {2006})}\BibitemShut
  {NoStop}%
\bibitem [{\citenamefont {Konoplya}\ and\ \citenamefont
  {Zhidenko}(2011)}]{konoplya_review_2011}%
  \BibitemOpen
  \bibfield  {author} {\bibinfo {author} {\bibfnamefont {R.~A.}\ \bibnamefont
  {Konoplya}}\ and\ \bibinfo {author} {\bibfnamefont {A.}~\bibnamefont
  {Zhidenko}},\ }\href {\doibase 10.1103/revmodphys.83.793} {\bibfield
  {journal} {\bibinfo  {journal} {Rev. Mod. Phys.}\ }\textbf {\bibinfo {volume}
  {83}},\ \bibinfo {pages} {793} (\bibinfo {year} {2011})}\BibitemShut
  {NoStop}%
\bibitem [{\citenamefont {Berti}\ \emph {et~al.}(2007)\citenamefont {Berti},
  \citenamefont {Cardoso}, \citenamefont {Gonzalez},\ and\ \citenamefont
  {Sperhake}}]{berti_PRD_2007}%
  \BibitemOpen
  \bibfield  {author} {\bibinfo {author} {\bibfnamefont {E.}~\bibnamefont
  {Berti}}, \bibinfo {author} {\bibfnamefont {V.}~\bibnamefont {Cardoso}},
  \bibinfo {author} {\bibfnamefont {J.~A.}\ \bibnamefont {Gonzalez}}, \ and\
  \bibinfo {author} {\bibfnamefont {U.}~\bibnamefont {Sperhake}},\ }\href
  {\doibase 10.1103/PhysRevD.75.124017} {\bibfield  {journal} {\bibinfo
  {journal} {Phys. Rev. D}\ }\textbf {\bibinfo {volume} {75}},\ \bibinfo
  {pages} {124017} (\bibinfo {year} {2007})}\BibitemShut {NoStop}%
\bibitem [{\citenamefont {Schutz}\ and\ \citenamefont
  {Will}(1985)}]{schutz_APJ_1985}%
  \BibitemOpen
  \bibfield  {author} {\bibinfo {author} {\bibfnamefont {B.~F.}\ \bibnamefont
  {Schutz}}\ and\ \bibinfo {author} {\bibfnamefont {C.~M.}\ \bibnamefont
  {Will}},\ }\href {\doibase 10.1086/184453} {\bibfield  {journal} {\bibinfo
  {journal} {Astrophys. J. Lett.}\ }\textbf {\bibinfo {volume} {291}},\
  \bibinfo {pages} {L33} (\bibinfo {year} {1985})}\BibitemShut {NoStop}%
\bibitem [{\citenamefont {Iyer}\ and\ \citenamefont
  {Will}(1987)}]{iyer_PRD_1987}%
  \BibitemOpen
  \bibfield  {author} {\bibinfo {author} {\bibfnamefont {S.}~\bibnamefont
  {Iyer}}\ and\ \bibinfo {author} {\bibfnamefont {C.~M.}\ \bibnamefont
  {Will}},\ }\href {\doibase 10.1103/physrevd.35.3621} {\bibfield  {journal}
  {\bibinfo  {journal} {Phys. Rev. D}\ }\textbf {\bibinfo {volume} {35}},\
  \bibinfo {pages} {3621} (\bibinfo {year} {1987})}\BibitemShut {NoStop}%
\bibitem [{\citenamefont {Matyjasek}\ and\ \citenamefont
  {Opala}(2017)}]{matyjasek_PRD_2017}%
  \BibitemOpen
  \bibfield  {author} {\bibinfo {author} {\bibfnamefont {J.}~\bibnamefont
  {Matyjasek}}\ and\ \bibinfo {author} {\bibfnamefont {M.}~\bibnamefont
  {Opala}},\ }\href {\doibase 10.1103/PhysRevD.96.024011} {\bibfield  {journal}
  {\bibinfo  {journal} {Phys. Rev. D}\ }\textbf {\bibinfo {volume} {96}},\
  \bibinfo {pages} {024011} (\bibinfo {year} {2017})}\BibitemShut {NoStop}%
\bibitem [{\citenamefont {Konoplya}\ \emph
  {et~al.}(2019{\natexlab{c}})\citenamefont {Konoplya}, \citenamefont
  {Zhidenko},\ and\ \citenamefont {Zinhailo}}]{konoplya_CQG_2019}%
  \BibitemOpen
  \bibfield  {author} {\bibinfo {author} {\bibfnamefont {R.~A.}\ \bibnamefont
  {Konoplya}}, \bibinfo {author} {\bibfnamefont {A.}~\bibnamefont {Zhidenko}},
  \ and\ \bibinfo {author} {\bibfnamefont {A.~F.}\ \bibnamefont {Zinhailo}},\
  }\href {\doibase 10.1088/1361-6382/ab2e25} {\bibfield  {journal} {\bibinfo
  {journal} {Class. Quantum Grav.}\ }\textbf {\bibinfo {volume} {36}},\
  \bibinfo {pages} {155002} (\bibinfo {year} {2019}{\natexlab{c}})}\BibitemShut
  {NoStop}%
\bibitem [{\citenamefont {Konoplya}(2003)}]{konoplya_PRD_2003}%
  \BibitemOpen
  \bibfield  {author} {\bibinfo {author} {\bibfnamefont {R.~A.}\ \bibnamefont
  {Konoplya}},\ }\href {\doibase 10.1103/PhysRevD.68.124017} {\bibfield
  {journal} {\bibinfo  {journal} {Phys. Rev. D}\ }\textbf {\bibinfo {volume}
  {68}},\ \bibinfo {pages} {124017} (\bibinfo {year} {2003})}\BibitemShut
  {NoStop}%
\bibitem [{\citenamefont {Blome}\ and\ \citenamefont
  {Mashhoon}(1984)}]{blome_PLA_1984}%
  \BibitemOpen
  \bibfield  {author} {\bibinfo {author} {\bibfnamefont {H.-J.}\ \bibnamefont
  {Blome}}\ and\ \bibinfo {author} {\bibfnamefont {B.}~\bibnamefont
  {Mashhoon}},\ }\href {\doibase https://doi.org/10.1016/0375-9601(84)90769-2}
  {\bibfield  {journal} {\bibinfo  {journal} {Phys. Lett. A}\ }\textbf
  {\bibinfo {volume} {100}},\ \bibinfo {pages} {231 } (\bibinfo {year}
  {1984})}\BibitemShut {NoStop}%
\bibitem [{\citenamefont {Chandrasekhar}\ and\ \citenamefont
  {Detweiler}(1975)}]{chandra_PRSL_1975}%
  \BibitemOpen
  \bibfield  {author} {\bibinfo {author} {\bibfnamefont {S.}~\bibnamefont
  {Chandrasekhar}}\ and\ \bibinfo {author} {\bibfnamefont {S.}~\bibnamefont
  {Detweiler}},\ }\href {\doibase 10.1098/rspa.1975.0112} {\bibfield  {journal}
  {\bibinfo  {journal} {Proc. R. Soc. A}\ }\textbf {\bibinfo {volume} {344}},\
  \bibinfo {pages} {441} (\bibinfo {year} {1975})}\BibitemShut {NoStop}%
\bibitem [{\citenamefont {Shu}\ and\ \citenamefont
  {Shen}(2005)}]{shu_PLB_2005}%
  \BibitemOpen
  \bibfield  {author} {\bibinfo {author} {\bibfnamefont {F.-W.}\ \bibnamefont
  {Shu}}\ and\ \bibinfo {author} {\bibfnamefont {Y.-G.}\ \bibnamefont {Shen}},\
  }\href {\doibase 10.1016/j.physletb.2005.05.077} {\bibfield  {journal}
  {\bibinfo  {journal} {Phys. Lett. B}\ }\textbf {\bibinfo {volume} {619}},\
  \bibinfo {pages} {340} (\bibinfo {year} {2005})}\BibitemShut {NoStop}%
\bibitem [{\citenamefont {D'Amico}\ and\ \citenamefont
  {Kaloper}(2020)}]{damico_PRD_2019}%
  \BibitemOpen
  \bibfield  {author} {\bibinfo {author} {\bibfnamefont {G.}~\bibnamefont
  {D'Amico}}\ and\ \bibinfo {author} {\bibfnamefont {N.}~\bibnamefont
  {Kaloper}},\ }\href {\doibase 10.1103/PhysRevD.102.044001} {\bibfield
  {journal} {\bibinfo  {journal} {Phys. Rev. D}\ }\textbf {\bibinfo {volume}
  {102}},\ \bibinfo {pages} {044001} (\bibinfo {year} {2020})}\BibitemShut
  {NoStop}%
\bibitem [{\citenamefont {Aneesh}\ \emph {et~al.}(2018)\citenamefont {Aneesh},
  \citenamefont {Bose},\ and\ \citenamefont {Kar}}]{kar_PRD_2018}%
  \BibitemOpen
  \bibfield  {author} {\bibinfo {author} {\bibfnamefont {S.}~\bibnamefont
  {Aneesh}}, \bibinfo {author} {\bibfnamefont {S.}~\bibnamefont {Bose}}, \ and\
  \bibinfo {author} {\bibfnamefont {S.}~\bibnamefont {Kar}},\ }\href {\doibase
  10.1103/PhysRevD.97.124004} {\bibfield  {journal} {\bibinfo  {journal} {Phys.
  Rev. D}\ }\textbf {\bibinfo {volume} {97}},\ \bibinfo {pages} {124004}
  (\bibinfo {year} {2018})}\BibitemShut {NoStop}%
\bibitem [{\citenamefont {Reed}\ and\ \citenamefont
  {Simon}(1975)}]{reed_simon_1975}%
  \BibitemOpen
  \bibfield  {author} {\bibinfo {author} {\bibfnamefont {M.}~\bibnamefont
  {Reed}}\ and\ \bibinfo {author} {\bibfnamefont {B.}~\bibnamefont {Simon}},\
  }\href@noop {} {\emph {\bibinfo {title} {Fourier Analysis,
  Self-Adjointness}}}\ (\bibinfo  {publisher} {Academic},\ \bibinfo {year}
  {1975})\BibitemShut {NoStop}%
\bibitem [{\citenamefont {Pal}\ and\ \citenamefont
  {Banerjee}(2016)}]{sridip_JMP_2016}%
  \BibitemOpen
  \bibfield  {author} {\bibinfo {author} {\bibfnamefont {S.}~\bibnamefont
  {Pal}}\ and\ \bibinfo {author} {\bibfnamefont {N.}~\bibnamefont {Banerjee}},\
  }\href {\doibase 10.1063/1.4972292} {\bibfield  {journal} {\bibinfo
  {journal} {J. Math. Phys.}\ }\textbf {\bibinfo {volume} {57}},\ \bibinfo
  {pages} {122502} (\bibinfo {year} {2016})}\BibitemShut {NoStop}%
\bibitem [{\citenamefont {Marple}(1986)}]{marple_book}%
  \BibitemOpen
  \bibfield  {author} {\bibinfo {author} {\bibfnamefont {S.~L.}\ \bibnamefont
  {Marple}},\ }\href@noop {} {\emph {\bibinfo {title} {Digital Spectral
  Analysis: With Applications}}}\ (\bibinfo  {publisher} {Prentice-Hall,
  Inc.},\ \bibinfo {address} {USA},\ \bibinfo {year} {1986})\BibitemShut
  {NoStop}%
\bibitem [{\citenamefont {{McDonough}}\ and\ \citenamefont
  {{Huggins}}(1968)}]{mcdonough_IEEE_1968}%
  \BibitemOpen
  \bibfield  {author} {\bibinfo {author} {\bibfnamefont {R.}~\bibnamefont
  {{McDonough}}}\ and\ \bibinfo {author} {\bibfnamefont {W.}~\bibnamefont
  {{Huggins}}},\ }\href {\doibase 10.1109/TAC.1968.1098950} {\bibfield
  {journal} {\bibinfo  {journal} {IEEE Trans. Autom. Control}\ }\textbf
  {\bibinfo {volume} {13}},\ \bibinfo {pages} {408} (\bibinfo {year}
  {1968})}\BibitemShut {NoStop}%
\bibitem [{\citenamefont {{Evans}}\ and\ \citenamefont
  {{Fischl}}(1973)}]{evans_IEEEAES_1973}%
  \BibitemOpen
  \bibfield  {author} {\bibinfo {author} {\bibfnamefont {A.}~\bibnamefont
  {{Evans}}}\ and\ \bibinfo {author} {\bibfnamefont {R.}~\bibnamefont
  {{Fischl}}},\ }\href {\doibase 10.1109/TAU.1973.1162433} {\bibfield
  {journal} {\bibinfo  {journal} {IEEE Trans. Audio Electroacoust.}\ }\textbf
  {\bibinfo {volume} {21}},\ \bibinfo {pages} {61} (\bibinfo {year}
  {1973})}\BibitemShut {NoStop}%
\bibitem [{\citenamefont {de~{Prony}}(1795)}]{prony}%
  \BibitemOpen
  \bibfield  {author} {\bibinfo {author} {\bibfnamefont {G.}~\bibnamefont
  {de~{Prony}}},\ }\href@noop {} {\bibfield  {journal} {\bibinfo  {journal} {J.
  l’\'{E}cole Polytechnique}\ }\textbf {\bibinfo {volume} {1}},\ \bibinfo
  {pages} {24} (\bibinfo {year} {1795})}\BibitemShut {NoStop}%
\end{thebibliography}%
\end{document}